%% file: main.tex
\documentclass[sigconf,nonacm=true]{acmart}

\usepackage{microtype}
\usepackage{graphicx}
\usepackage{subcaption}
\usepackage{booktabs} %

\usepackage{balance}

\usepackage{hyperref}

\usepackage{times}
\usepackage{color}
\usepackage{xspace}
\usepackage{amsmath}
\usepackage{url}
\usepackage{enumitem}

\def\ie{\emph{i.e.}}
\def\eg{\emph{e.g.}}

\newcommand{\system}{{FLaaS}\xspace}

\begin{document}

\title{\system: Cross-App On-device Federated Learning in Mobile Environments}

\author{Kleomenis Katevas}
\affiliation{
    \institution{Brave Software}
    \country{London, UK}
}
\author{Diego Perino}
\affiliation{
    \institution{Telef\'onica Research}
    \country{Barcelona, Spain}
}
\author{Nicolas Kourtellis}
\affiliation{
    \institution{Telef\'onica Research}
    \country{Barcelona, Spain}
}

\input{sections/00_abstract.tex}
\maketitle
\input{sections/01_introduction}

\input{sections/03_overview}

\input{sections/04a_backend}

\input{sections/04b_clients}

\input{sections/06_evaluation}

\input{sections/02_background}

\input{sections/10_conclusions}

\section*{Acknowledgements}

The research leading to these results received partial funding from the EU H2020 Research and Innovation programme under grant agreements No 830927 (Concordia), No 871793 (Accordion), No 871370 (Pimcity), and No 101021808 (SPATIAL).
These results reflect only the authors' view and the EU Commission are not responsible for any use that may be made of the information it contains.

\balance
\bibliography{references}
\bibliographystyle{acm}

\end{document}

%% file: sections/00_abstract.tex
\begin{abstract}
Federated Learning (FL) has recently emerged as a popular solution to distributedly train a model on user devices improving user privacy and system scalability.
Major Internet companies have deployed FL in their applications for specific use cases (\eg,~keyboard prediction or acoustic keyword trigger), and the research community has devoted significant attention to improving different aspects of FL (\eg,~accuracy, privacy, efficiency).
However, there is still a lack of a practical system to enable easy collaborative cross-silo FL training, in the context of mobile environments.
In this work, we bridge this gap and propose \system, an end-to-end system (\ie,~client-side framework and libraries, and central server) to enable intra- and inter-app training on mobile devices with different types of IID and NonIID data distributions, in a secure and easy to deploy fashion.
Our design solves major technical challenges such as on-device training, secure and private single and cross-app model training, while being offered in an ``as a service'' model.
We implement \system for Android devices and experimentally evaluate its performance in-lab and in-wild, on more than 140 users for over a month.
Our results show the feasibility and benefits of the design in a realistic mobile context and provide several insights to the FL community on the practicality and usage of FL in the wild. 
\end{abstract}

%% file: sections/01_introduction.tex
\section{Introduction}

Federated Learning (FL)~\cite{konecny2016federated-learning} enables data owners to securely train a Machine Learning (ML) model between them, by sharing only private models trained locally on their data.
FL, coupled with Differential Privacy (DP)~\cite{geyer2017FL-DP-central}, Trust Execution Environments (TEE)~\cite{ppfl}, or other methods, has been adopted by major tech corporations (\eg,~Google~\cite{bonawitz2019FL-sysml}, Apple~\cite{apple-fl}, Meta~\cite{huba2022papaya-mlsys}, etc.) due to its properties such as scalability, privacy-preserving (PP) data modeling and performance.

Typical FL use cases currently employed by industry focus on building an ML model for a specific ML task by applying FL across user devices, \ie,~\textit{cross-device}, on user local data (\eg,~GBoard~\cite{bonawitz2019FL-sysml}).
These deployments are based on proprietary code, embedded in owner products, and typically designed on a per application (app) basis.
This clearly limits the usage of FL to a few big players able to sustain cost and risk of developing and using FL.
To change this status quo, recent start-up or open source efforts (\eg,~\cite{openmined2020, datafleets2020, DML2018, integrate-ai, sherpa-ai}) aim to provide FL tools for B2B customers or end-users, where FL is performed across company servers, \ie, \textit{cross-silo}, on cloud-collected data (\eg,~between health providers, or financial institutions, etc.).
Furthermore,~\cite{flaas} even advocates for an FL-as-a-Service system in mobile settings.

Despite these early efforts, there are several novel scenarios where FL can be applied that remain unexplored, such as the \textit{cross-app} mode: a combination of \textit{cross-silo} and \textit{cross-device} for an ML task that apps share.
For example, two e-health apps (\eg,~fitness and nutrition apps) that collect partially similar data, want to collaborate and solve the same ML problem with better accuracy (\eg,~detecting when their user is at risk of diabetes).
If the apps could locally share their data or models in a PP fashion, they could train a unique FL model across their users' devices that helps both apps, while not violating data owner's privacy.
Under the \textit{cross-app} mode, even multi-task ML problems~\cite{marfoq2021multi-taskFL, corinzia2021multi-taskFL} could be collaboratively solved between apps.

Unfortunately, many system challenges remain to be solved to materialize such scenarios in the mobile context, and even democratizing FL to small and medium companies in as-a-service fashion.
First, there is lack of libraries supporting background ML training on mobile devices, and especially for collaborative, cross-app scenarios.
Second, there is no support for cross-app local sharing of data or models for collaborative training: it requires secure and PP mechanisms for storage, communication, and permissions management across apps.
Finally, existing FL methods do not provide an easy to use model for app developers based on simple APIs and tools as ML-as-a-Service counterparts (\eg,~Amazon Web Services~\cite{aws-mlaas2020}, Google Cloud~\cite{googlecloud-mlaas2020} or Microsoft Azure~\cite{azure-mlaas2020}).

\begin{figure}
    \centering
    \includegraphics[width=\linewidth]{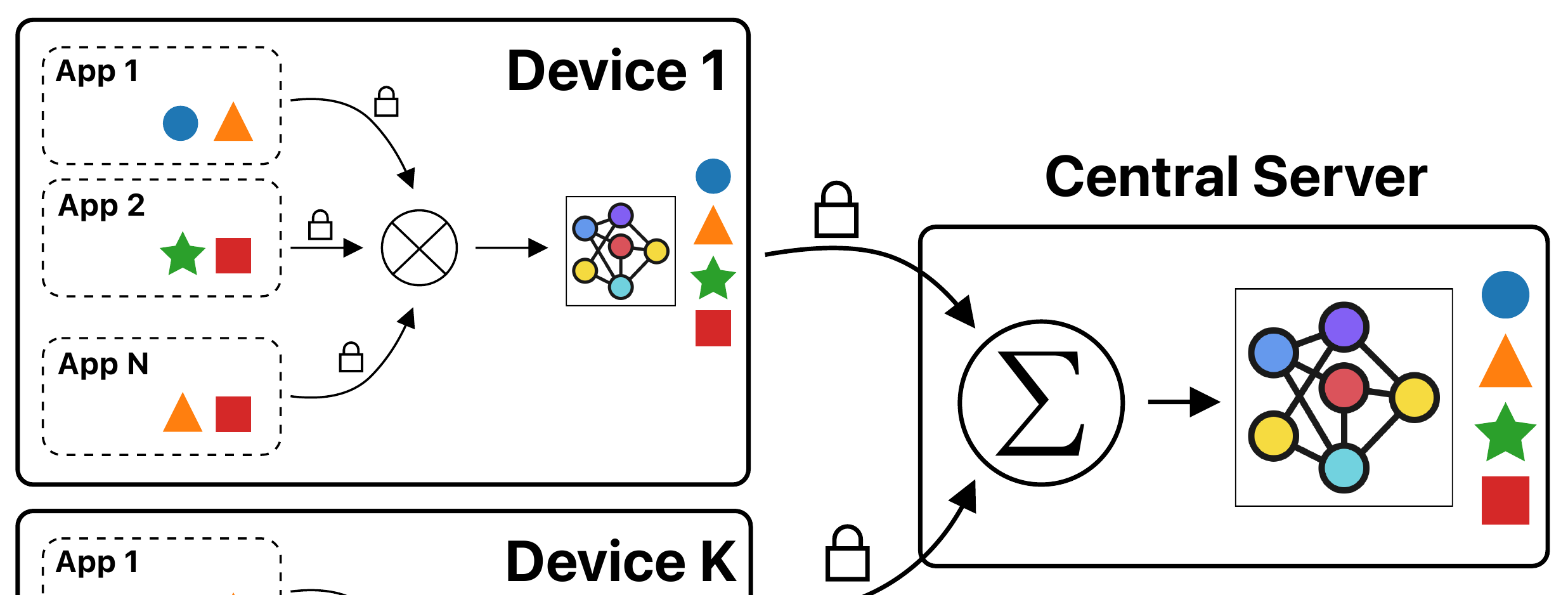}
    \caption{\system Overview: A set of K devices with N installed apps securely share their heterogeneous samples or locally trained models with the local service. They are aggregated locally (denoted by $\otimes$) in order to produce a local model. The model parameters from all K devices are securely aggregated by applying FedAvg (denoted by $\Sigma$) following the traditional federated learning approach.}
\vspace{-0.45cm}
    \label{fig:introduction-figure}
\end{figure}

In this paper, we focus on solving these challenges and present \system, the first practical FL framework for mobile environments that enables \textit{cross-app} FL, and \textit{as-a-Service} (Figure~\ref{fig:introduction-figure}).
Our design includes a client-side middleware and library enabling third-party mobile apps to train FL models on-device, either independently or in a joint fashion. We investigate two types of cross-app FL modeling: shared samples and shared models.
Cross-app training is based on a set of secure and PP communication and storage primitives, while the library makes seamless the integration of FL into existing apps.
\system also includes a centralized component to automatically orchestrate the FL process, with an easy interface for app developers.
Our design is end-to-end and includes several system optimizations required to deploy it in the wild in a scalable and robust fashion.

We implement \system for Android-based mobile devices (\ie,~client-side elements), while leveraging a popular cloud-based platform for the centralized component.
The complete source-code of our system is available at: \url{https://github.com/FLaaSResearch}.
Finally, we evaluate \system feasibility, overhead and performance via in-lab and real-world experiments with more than 140 real users.
The former allows us to evaluate the cost of the system under controlled settings: we observe devices spend a limited amount of time (\ie,~few minutes) and resources (\ie,~energy and CPU) in training of an ML model during an FL round, and communication with the system.
The evaluation with real users allows us to analyze the ML performance and cost in real settings, as well as derive critical insights for any FL system for mobile environments in the wild.
We observe that even if a few 10s of devices actually report trained models, \system builds a global FL model that compares in utility with a centralized model, while having very limited impact on user experience.
We also observe that cross-app modeling with shared samples achieves, on average, $24.10\%$ higher accuracy across all data distributions when compared with the shared models approach, and this difference is pronounced in 3 diverse and realistic NonIID data distributions tested across participating users.
To the best of our knowledge this is the first work to design and experimentally evaluate a practical, FL-as-a-Service system for collaborative cross-mobile app ML training in the wild.

%% file: sections/03_overview.tex
\section{System Overview}
\label{sec:principles}

\begin{figure*}[t]
    \centering
     \includegraphics[width=\linewidth]{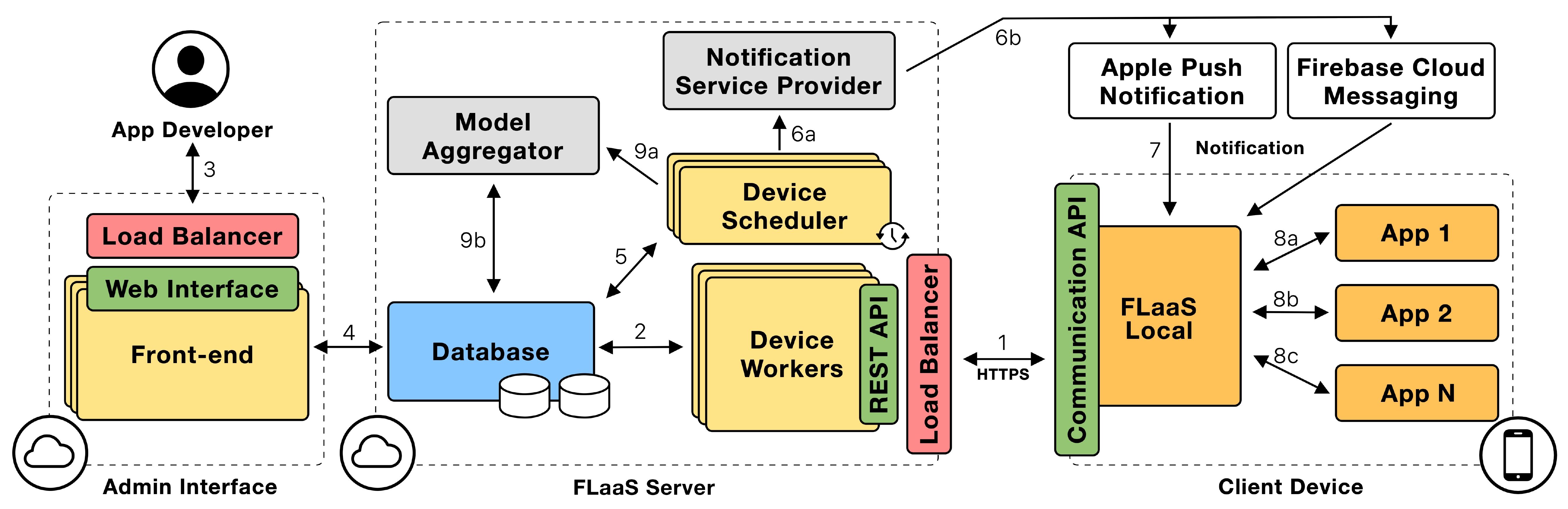}
    \vspace{-0.8cm}
    \caption{Overview of \system ML modeling architecture. We assume a set of client devices participating in \system device infrastructure, periodically reporting their status through the backend's REST API (1) to the DB for logging (2).
    An app developer uses the Admin Interface to create a new \system FL project (3), whose configuration is stored in the DB (4).
    The Device Scheduler detects the new \system project (5), queries the device statuses from the DB (5), and sends an FL training request using its Notification Service provider (6a) to external services such as APN and FCM (6b).
    Each device's \system Local receives the FL training request (7) and requests from the participating third-party apps to receive either their local samples for training, or to train their own models (8a, b and c).
    It then conducts the necessary on-device model training (if it received samples) or model aggregation and averaging depending on the training mode. When a substantial number of reported models is received by the Server (1), the Device Scheduler instructs the Model Aggregator to accumulate the received models (9a and 9b), marks the FL round as complete, and continues with the next FL round until the project is complete (\ie,~stopping requirements are reached).
    }
    \label{fig:architecture}
\vspace{-0.4cm}
\end{figure*}

In this section, we discuss 5 main challenges that \system addresses with its design, in providing a practical end-to-end FL system for mobile apps.

\subsection{Design Challenges}
\noindent
\textbf{1. Easy to use.}
\system needs to be 1) easy to incorporate in existing apps; 2) provide a simple interface to configure the FL training process.
For the first issue, we design two on-device components: a service and a library.
The service is to be installed in devices participating in the FL process.
The library is to be integrated in existing apps and can be used by calling its APIs, thus, requiring adding only a few lines of code.
For the second issue, we provide a user interface (UI) similar to the ones provided by existing MLaaS.

\noindent
\textbf{2. Multi use case coverage.}
\system should be able to support at least two types of ML modeling scenarios:\\
\noindent
\textit{i) Single-app FL modeling:} Typically, a specific app is interested in building a ML model to solve a specific task: \eg,~a nutrition app that wants to model its users' behavior and preferences to provide better food recommendations, and thus wants to utilize the local data per user/device.\\
\noindent
\textit{ii) Cross-app FL modeling:} A group of two or more apps, concurrently installed on the same device, want to collaborate and build a shared model to solve a task, identical for each app, but on more, shared and possibly heterogeneous data.
For example, a fitness app and a nutrition app may want to build a joint ML model for better health recommendations, to push their users away from diabetes risk, based on data coming from each app's local store.

\noindent
\textbf{3. On-device training}.
Currently there are no libraries enabling background ML training, and between mobile apps, to support our envisioned use cases.
Thus, we design an on-device, background ML training module building on TensorFlow Lite framework that we extend to be integrated, and support the different functionalities of \system, and to provide the different training modes, \ie,~single-app and cross-app.

\noindent
\textbf{4. Secure and private training.}
\system needs to enable a secure and private training process. 
First, we enable a secure communication channel between apps embedding the library and the \system service supporting different types of data, \ie,~numerical, text and binary.
Second, data and model storage on device are secured using per app and service spaces with permissions specified by developers and enforced using an on-device data management engine.
Finally, we need to guarantee secure authentication and communication between the \system back-end and the training client devices.
This is achieved via authentication tokens for apps and \system service, and using state of art encrypted communication mechanisms. 
We assume developers and the system are trusted, and that attacks to the training process or models, such as membership/attribute/data inference attacks~\cite{shokri2017membership, nasr2019comprehensive, sablayrolles2019white} can be solved by integrating in our design solutions with TEEs such as~\cite{ppfl}, or by injecting DP noise during the FL process~\cite{dwork2014algorithmic}.

\noindent
\textbf{5. Production ready.}
\system needs to be deployable in the wild and to deal with production environment challenges such as scalability and robustness.
To this goal, we leverage several system and software engineering optimizations both at the back-end (\eg,~easy scalability for load balancing, notification service, etc.), and device (\eg,~opportunistic ML training based on available battery, optimized data storage and inter-app communication for joint-app modeling).

\subsection{Main Components}
To address these challenges, we propose \system.
Figure~\ref{fig:architecture} provides an overview of the system architecture, along with the list of steps required to perform an FL training process.
The 4 core system elements (detailed in Sec.~\ref{sec:design}\&~\ref{sec:client-devices}) are:
\begin{itemize}[leftmargin=10pt, noitemsep,topsep=0pt]
    \item \emph{Admin Interface} (Backend,\textsection~\ref{sec:admin-interface}): The main front-end interface for app developers, which is used to bootstrap, configure, execute and monitor an FL project within \system.
    \item \emph{Server} (Backend,\textsection~\ref{sec:server}): A scalable, cloud-hosted service orchestrating and monitoring the FL process on the app developer's behalf.
    \item \emph{Notification Services} (Backend,\textsection~\ref{sec:notification-service}): Mobile push notification services used for push communications to client devices for new FL projects and rounds execution: typically Apple Push Notification (APNs)~\cite{APNs} and Firebase Cloud Messaging (FCM)~\cite{FCM} are used in the mobile context.
    \item \emph{Client Devices} (\textsection~\ref{sec:client-devices}): The set of mobile devices that could execute the FL tasks on the apps' local data.

\end{itemize}

We elaborate on these elements in the next two sections.

%% file: sections/04a_backend.tex
\section{\system Back-end Design}
\label{sec:design}

\subsection{Admin Interface}
\label{sec:admin-interface}

This is the administrator UI for developers to bootstrap, configure, monitor, and terminate an FL project executed within \system.
Developers can use the UI for registering their apps with the system, produce the authentication tokens that will be used in their app and configure the data and model access policies to be enforced by \system at device level.
Thus, they can create new FL projects to be executed from \system-registered devices, make a pre-selection of participating apps and devices, and initiate the FL training process.
For details on device registration with \system see \textsection~\ref{sec:client-devices}.

Each app developer interested in running an FL project using \system can configure their project with a number of available system, ML and FL parameters:
\begin{itemize}[leftmargin=10pt,noitemsep,topsep=0pt]

    \item \texttt{ML model architecture:} A list of available ML models that can be used in the on-device ML training task.
    New models can also be uploaded by the developer, as long as they are supported by the on-device ML training engine.

    \item \texttt{ML model training mode:} 1) \emph{Single-app (SA)} (\ie, a model per app); 2) \emph{Joint Samples (JS)} (\ie,~data shared between apps); 3) \emph{Joint Models (JM)} (\ie,~model trained on app's local data, then model shared between apps, more:\textsection\ref{sec:client-devices}).

    \item \texttt{ML model training epochs:} Number of epochs (or iterations) the SGD will be executed on the data of an app, while training the model, during a given FL round.

    \item \texttt{Policy Management:} Depending on model training mode selected, different policy options are available to the developer: how data or model parameters would be shared between which apps, for how long, etc. More: \textsection~\ref{sec:client-devices}.

    \item \texttt{FL training rounds}*: Number of iterations of the FL process (\ie,~global model sent to clients, trained locally, and returned to server for aggregation).

    \item \texttt{FL training round duration}*: Maximum time duration that an FL training round can last.

    \item \texttt{FL model loss}*: Minimum value of optimization function loss to be achieved by the aggregated FL model.

    \item \texttt{Device power availability:} Power of devices that should participate in the FL project: only devices power-plugged AND/OR devices battery-powered, with battery level $>X\%$ (if defined).
    Lower $X\%$ allows more devices to participate.

    \item \texttt{Number of available devices:} Number of devices that need to be reached based on the device power availability setting, for an FL training round start request to be sent to available and responded devices.

    \item \texttt{Number of model parameters received:} Number of ML model parameters that should be successfully received for an FL training round to be considered valid.
\end{itemize}
*:~Can be used as a stopping criterion for the FL round.

When the developer defines a new project using the UI (step 3, Fig.~\ref{fig:architecture}), the project is pushed to the \emph{\system Server} (step 4) for (i)~storing its configuration, (ii)~executing the project in the available client devices, (iii)~monitoring the project's health and progress, (iv)~terminating the project when stopping criteria are reached.
The Admin UI is highly scalable using a Load Balancer to accommodate concurrent developers managing their \system projects.

\subsection{\system Server}
\label{sec:server}

It is a cloud-hosted service in charge of executing the FL project previously created at the Admin Interface, while coordinating the Client Devices.
It consists of the following modules:
1) Database, 2) Device Schedulers, 3) Model Aggregator, 4) Device Workers.
Next, we explain each.

\subsubsection{Database (DB)}
The system requires two separate databases responsible for storing and querying data: a \emph{Relational DB} and an \emph{Object Storage DB}.
On the one hand, the \emph{Relational DB} is an SQL-based database for hosting project-related metadata.
These include the FL project configuration as defined by the developer through the Admin Interface, information reported by the devices (\eg,~power availability, status, step 1+2), authentication credentials, project status (training or waiting), etc.
On the other hand, \emph{Object Storage DB} is a file storage for hosting files.
These include the ML model parameters received by Client Devices, ML performance results, ML performance plots created and to be shown to the developer via the Admin Interface, etc.

\subsubsection{Device Schedulers}
\label{sec:device-schedulers}
This is a set of workers, one per FL project, configured to be executed every $T$ minutes.
$T$ can be set by the system administrator depending on the FL project load, device availability and FL project requirements.
The role of this module is to verify the status of all active FL projects, which can either be at \emph{waiting} or \emph{training} mode, and execute the related management tasks per project.
Each new FL project has an initial state of \emph{waiting} and for the scheduler there are two possible conditions:

\noindent
\textbf{1. FL project is in \emph{waiting} state.} In this case, the scheduler executes a query on the Relational DB to identify how many devices meet the \texttt{device power availability} criterion defined by the developer via the Admin Interface. This query also returns the list of devices available across the system that have the desired power status (battery of at least $X\%$ and/or plugged-in), for an FL training round to be performed for the specific FL project. If the criterion is reached, a new FL round can start: the specific scheduler changes the state of the FL project to \emph{training}, stores this information to the Relational DB for future reference and sends a \emph{Device FL Training Request} through the \emph{Notification Service} to the devices identified as available (see next paragraph for details).

\noindent
\textbf{2. FL project is in \emph{training} state.} In this case, the \emph{Device Scheduler} checks how many model training responses (model training parameters and on-device ML performance results) have been received from devices.
If the FL round is \emph{completed} (\ie,~enough models were received), the \emph{Device Scheduler} requests from the \emph{Model Aggregator} to apply Federated Aggregation (FedAvg)~\cite{mcmahan2017-FL} to the parameters of received models to generate a new global FL model (see details next), and sets the project in \emph{waiting} state to begin a new round.

An alternative option supported by the system is to wait for \texttt{FL training round duration} expiration before it performs FedAvg and subsequent steps: this would potentially collect more models from clients thus improving accuracy, but would extend the round duration to its maximum.
If this duration is reached and there are not enough ML models received (as defined earlier), the specific FL round is marked as \emph{invalid}, and the project is set to \emph{waiting} state to begin a new FL round.

This iterative process is followed until the requirement of number of FL rounds is fulfilled.
Alternatively, the developer may also define a desired \texttt{FL model loss} for the project to be reached, before it is completed.
If this parameter is set, at the end of each round, the scheduler also requires from the \emph{Model Aggregator} to assess the global ML model's performance such as test accuracy (evaluated at the Server) and optimization function loss (as reported by the Client Devices), and persists them on the Object Storage DB.
Thus, the Scheduler continues to initialize FL training rounds for the specific project until any of the desired model loss or rounds are reached.

\subsubsection{Model Aggregator}
This component requests from the \emph{Object Storage DB} for all ML model parameters (weights) reported by devices for a specific FL project and FL round.
Then, it accumulates and applies the FedAvg algorithm in order to produce the global FL model of the next (or final) FL round of the given project.
Optionally, Global $DP$ applied in a centralized fashion~\cite{geyer2017FL-DP-central} can be enforced to the produced model for protecting the users' privacy from relevant model attacks, such as inference attacks~\cite{hitaj2017deep, melis2019exploiting} and other types of attacks~\cite{hitaj2017deep, melis2019exploiting}.

\subsubsection{Device Workers}
This module contains a series of containerized workers
responsible for interacting with the client devices for FL model training assigning and responses received.
Each worker includes a dedicated REST-based communication API that allows the interaction of the worker with the client devices involved in a particular FL project.
Such interaction is initiated by the client device (\eg,~to report its device status).
However, it can also be indirectly triggered by the Server through a push notification (\eg,~a training request) to the device, that in turn contacts the Server. 
Which worker to handle what FL project is decided by the system Load Balancer, as explained in the next paragraph.
Each worker includes the following REST endpoints, all exposed to the client devices using a token-based authentication:
\begin{itemize}[leftmargin=10pt,noitemsep,topsep=0pt]
    \item \texttt{Report Availability}: Device status (\eg,~battery level, charging state, memory available, etc.) reported to the system and used by the Device Scheduler to find opportune moments for requesting device training tasks.
    \item \texttt{Get model}: Download the model's parameters of a particular round of a project.
    \item \texttt{Join Round}: Declares that a device has received the training request and will execute the requested training task of a particular round.
    \item \texttt{Submit Model}: Upload the parameters of a trained model to the server.
    \item \texttt{Submit Results}: Upload training results (\ie,~loss, device performance during training, etc.) to the server.
\end{itemize}

\subsubsection{Load Balancer}
This is the entry point of the Server when interacting with the Client Devices.
It automatically distributes all HTTP(S) requests sent to the Server's hostname(s) to the Device Workers.
It also monitors the performance of the Device Workers and automatically creates new instances when required (\ie,~when the Server is under heavy network traffic load due to several, simultaneous devices reporting status, models, results, etc.).

\subsection{Notification Service}
\label{sec:notification-service}

The Notification Service (NS) is used for secure, asynchronous, and one-way communication from the Server to Client Devices.
While it can generally depend on various types of technologies (\eg,~message brokers, TCP connections, etc.), in the context of mobile computing it takes the form of a Push NS.
These are cloud-based, highly efficient services for iOS and Android to propagate information to a previously registered mobile app, either in a form of a visual popup, but also invisible (silent) notifications that deliver JSON structured data to an app.
They use secure TCP communication to directly push data to a device's registered app through dedicated cloud services provided by Apple (APNs)~\cite{APNs} and Google (FCM)~\cite{FCM}.
To utilize such services, a NS Provider is required: either custom made or an existing one (\eg,~Amazon SNS, Pushwoosh, etc.).
Silent notifications (\ie,~notifications that do not appear to the user but wake up the app in the background) are usually received with an unspecified delay from the app.

In our system, such communication is used for sending FL training requests to the participating devices.
Each training request sent to devices consists of the following fields:
\begin{itemize}[leftmargin=10pt,noitemsep,topsep=0pt]
    \item \texttt{Request ID}: A unique ID per request.
    \item \texttt{Expiration Date}: The date after which the request will be ignored from the participating device.
    \item \texttt{Project ID}: A unique ID of the FL project that the ML training will be conducted.
    \item \texttt{FL round}: The round for which this request is relevant.
    \item \texttt{ML training mode}: A selection between Joint Samples (JS) or Joint Models (JM), as explained in \textsection~\ref{sec:client-devices}.
\end{itemize}

As it is apparent from the information included in each notification, these messages from the Notification Service are used to alert the devices of pending FL training tasks in the system, that they can contribute to if available.

\subsection{Back-end Implementation}
\label{sec:backend-implementation}

The Admin Interface and Server are hosted in Heroku Cloud Application platform~\cite{heroku}, under the EU region.
The platform uses Linux-based shared containers called \emph{Dynos} to enable easy deployment and scaling of the system modules.
In fact, we use Standard-1X Dynos (512MB RAM) and leverage platform load balancing and scheduling functions.
The NS Provider is hosted in a separate organization domain (Pushwoosh~\cite{pushwoosh}).
Overall, the back-end is implemented in 1.6k lines of Python code.

\noindent
\textbf{Admin Interface:}
This module is hosted on Heroku using the gunicorn web server v20.1.0.
An instantiation of the Web interface is executed inside a Web Dyno container.
A Heroku Load Balancer allows scaling to more Dynos to accommodate concurrent developers interacting with \system.

\noindent
\textbf{Databases:}
All data are stored into two separate databases, as explained earlier.
A PostgreSQL v13.4 instance is used for the Relational DB, and an Amazon S3 Data Store is used as an Object Storage DB.

\noindent
\textbf{Device Schedulers:}
Each scheduler is implemented within a separate one-off (worker) Dyno that is only instantiated periodically every $T$ minutes, as defined earlier.
This periodic launching is handled by the \emph{Heroku Scheduler service}, by starting a new Dyno (one per FL project) to perform its assigned tasks:
1) start a new or pending FL round (if device availability requirement is met),
2) stop an earlier FL round if completion criteria have been met,
3) launch the Model Aggregator for preparation of new global model,
4) terminate the FL project if completion criteria have been met.

\noindent
\textbf{Device Workers:}
Each Worker is implemented within a separate, continuously running Web Dyno.
All HTTPS traffic is handled by a Load Balancer instantiating new Dynos to handle elevated incoming workload from Devices.
Each Device Worker uses Django v3.2 under Python v3.8.12 and has a dedicated REST API based on Django REST framework v3.12.4.
Communication between Device Workers and Client Devices is encrypted using TLS v1.3 over HTTPS.

\noindent
\textbf{Notification Service Provider:}
We use the Pushwoosh push notification services~\cite{pushwoosh} configured with both FCM (for Android devices) and APN (for future extension to iOS).

%% file: sections/04b_clients.tex
\section{\system Client Devices}
\label{sec:client-devices}

\begin{figure}
    \centering
    \includegraphics[width=\linewidth]{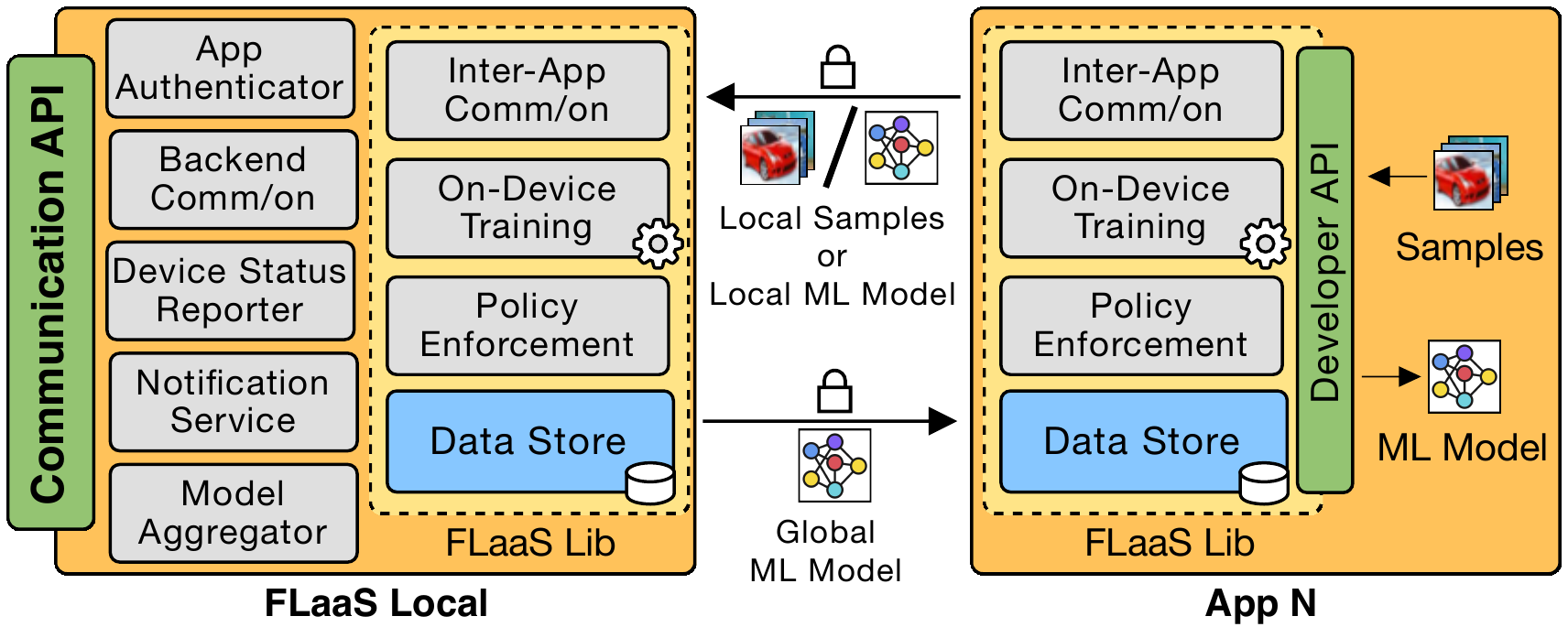}
    \vspace{-0.8cm}
    \caption{Inter and intra-app modules on Client Devices}
\vspace{-0.45cm}
    \label{fig:client-device}
\end{figure}

The most important piece of \system is the set of client devices in charge of computing the ML training tasks orchestrated by the Server for the execution of an FL project.
On such devices, there are two main modules: Local and Library (Lib).
These modules facilitate on-device functionalities (Figure~\ref{fig:client-device}): App authentication, inter-app communication, access policy management, ML model training, data storing, and status reporting.
Each function is elaborated next.

\subsection{\system~Local}
This is a standalone service that needs to be installed on the device and provides the core FL functionality to the device in accordance with requests sent by the Server.
We do not make any assumption on how this service is installed.
We envision this service can be pre-installed on the device by the device or OS manufacturer or provider. 
Alternatively, an interested third-party app developer can recruit a device owner to install its app with proper user incentives (\eg,~promise of better app experience due to ML model personalized on user data).

Local provides authentication and secure communication of the device to the Server, and periodically reports the device status (akin to a heartbeat), with different details such as battery level, charging state, connectivity state, to the Server for consideration.
The set of statuses from all participating client devices is in fact the one collected and analyzed by the Server to decide which devices to invite for upcoming FL training tasks (as explained in \textsection~\ref{sec:device-schedulers}).

Local is also the module that receives messages from the \emph{Notification Service} with configurations for pending FL project tasks related to specific third-party apps.
These task configurations are then communicated to the appropriate apps for execution, within their Library (see next).

\subsection{\system Library}
This is a mobile app library integrated within each \system-enabled third-party app willing to participate in FL training processes.
This library has a set of simple APIs that are called by the developer in the app code.
More particularly, the API consists of calls to register the app's secure token (created through the Admin Interface) and to provide controlled (secured through policies) access to data the app is willing to share.
Under the hood, it implements a set of functions that allows the app to securely communicate with Local to receive ML modeling configurations based on FL project of relevance to the app, execute the appropriate on-device ML training, and share data or model parameters with Local.

\subsection{\system Security}
\label{sec:device-security}
Our system depends on two communication channels:

\noindent \textbf{Back-end communication} of Server with each user's Local.
Security, integrity, and confidentiality are automatically enforced using secure $TLS$ communication.
Device authentication is implemented using a set of pre-installed authentication tokens, saved within the device's keystore system (\ie,~Android keystore or iOS keychain).

\noindent \textbf{Inter-app communication} between Local and third-party apps.
An app authenticates with the Server via its Library through the Local, to enable inter-app communication and collaborative ML model training.
We use a symmetric key approach~\cite{yassein2017symmetric}, based on authentication tokens produced by the Developer through the Admin Interface and also securely stored to the device's keystore system.
Each token has an expiration date but can also be invalidated by Local at any moment a developer needs to.
All communication takes place within the device and can support encrypted numerical, text and binary types of data.
The purpose of this channel, materialized via the Library instantiated in each app and Local, is to enable communication between Local and apps, and communicate project-related settings, get training samples, or trained models, etc.
In the future, we will explore advanced cryptographic methods using public key cryptography with external certification authority. %

\subsection{On-device ML Training}
We assume a set of apps, $i \in \mathbb{N}$, installed on device $k \in \mathbb{K}$, are interested in building $FL$ models with \system.
There are two types of ML training supported by \system and executed on the client devices: Single-app and Cross-app FL modeling.
Both options require ML modeling at the app or at Local level.
This is performed by an on-device ML training engine that exists in the Library.

\noindent \textbf{1. Single-app FL modeling.}
In this case, each app $i$ of device $k$ trains a model on its own data only.
The starting configuration for this training is received by the Library of the specific app involved, and includes the base model to be trained, access policy requirements on the local data to be used (see details next), and model hyperparameters.
If the developer requires DP-noise to be injected, this is done at this ML training stage.

After local training is done, each app sends to Local its model built: $F^i_k(w), \forall i \in \mathbb{N}$.
When Local has all such models from unique apps, it compresses and transmits them to \system Global.
Global performs FedAVG across all reported local models of app $i$ (i.e., $F^i_k(w), \forall k \in \mathbb{K}$) and builds one global weighted average model per app, which it then communicates back to the participating devices per app:
$f^i(w), \forall i \in \mathbb{N}$.
Finally, Local distributes this global model to each app for next FL round.

\noindent
\textbf{2. Cross-app FL Modeling.}
A group $g$ of third-party apps (e.g., $g$=$\{i,j\}$, $i,j$ $\in$ $\mathbb{N}$), interested in collaborating to build a joint model that solves the same ML task (\eg,~object recognition on images from both Google Photos and Gallery apps) have two \system \texttt{ML training modes}: JS and JM.

\emph{2a. Joint Samples (JS) mode:} The apps in $g$ share training samples with Local, appropriately packaged to a format pre-agreed by the developers.
DP-noise may be added by each app, in a pre-training stage.
Local checks the received samples from each app for formatting, and then merges them before performing ML model training on all.
If these third-party apps require it, Local may also inject further DP-noise during the ML training.
Then, Local communicates with the Server the resulting ML model as in the Single-App mode.

\emph{2b. Joint Models (JM) mode:} The apps in $g$ conduct the ML model training individually using their local samples, as in the Single-app mode (with or without DP-noise).
Then, each app shares only its ML model parameters with Local, which then applies FedAvg to the received model and produces a joint model.
It finally communicates to the Server the joint ML trained model across apps in $g$ of device $k$: $F^g_k(w)$.
Global performs FedAVG across all collected local models and builds a global weighted averaged model, for each collaborating group $g \in \mathbb{G}$ of apps.
Then, it communicates each such global model back to the participating devices: $f^g(w), \forall g$.
Finally, each Local distributes the global model to each app of the collaborating group $g$, $f^g(w), \forall i \in g$.

\subsection{On-device data management}

To support these two training modes, Local and Library are responsible for the following data management functions:

\noindent
\textbf{Access Policy Management:} A policy enforcement engine (\eg,~\cite{bagdasaryan2019ancile}) defined by the developer in the Admin Interface.
The developer can define:
\begin{itemize}[leftmargin=10pt,noitemsep,topsep=0pt]
    \item Which data (for $JS$) or ML models (for $JM$) to share with Local for the Cross-app training.
    \item What type of pre-processing (\eg,~data filtering) (for $JS$) should be applied before sharing.
    \item How much DP-noise should be added on samples before sharing (in case of $JS$), or while training the model (in case of $JM$) (following best practices in FL with DP~\cite{geyer2017FL-DP-central}).
    \item Which other apps can receive these shared data or ML models (for now, only Local can receive them, but future training modes could require sharing data and/or models with other apps).
    \item How many times to allow this sharing between apps: a) only once, b) $N$ times (\ie,~for $N$ FL rounds), c) until certain deadline is reached.
    \item Minimum loss of locally trained model (in case of $JM$).
\end{itemize}

\noindent
\textbf{Data Store:} This is a secure data store of Local and Library (one for each involved app).
Local can temporarily host binary data such as training samples sent by apps and prepared for on-device training by Local, or ML weights sent by apps and prepared for model averaging by Local.
Each app's Library can host training weights sent by Local and used as basis for in-app ML training.
Each datum has an expiration condition and automatically gets deleted when the condition is met (\eg,~trained model is sent).

\subsection{Client Device Implementation}
\label{sec:client-devices-implementation}

\noindent
\textbf{Supported Mobile Operating System:}
\system currently supports Android-based ($\geq$8.0) devices, since iOS does not provide (yet) in-app communication features necessary for instructing third-party apps to perform on-device cross-app ML training, either using shared data or shared models.

\noindent
\textbf{Device Authentication:}
Device authentication with the back-end is implemented using JWT with djangorestframework-simplejwt v5.0.0.
The framework is configured to produce tokens expiring after 10 minutes but can be refreshed automatically by the client using \emph{refresh tokens} function, with rotating expiration time of 1 day.

\noindent
\textbf{\system Library:}
Third-party apps participating in projects implement this library to:
receive configuration setups and perform ML tasks in single-app or JM modes, send samples to Local for JS mode, or
communicate model parameters to Local for FedAvg.
The Library is implemented with 1.5k lines of Java code, with:

\begin{itemize}[leftmargin=10pt,noitemsep,topsep=0pt]
    \item TensorFlow Lite v2.6.0: Deep learning framework used for on-device ML training.
    \item WorkManager v2.7.0: Android framework for scheduling deferrable, asynchronous tasks within apps.
\end{itemize}

\noindent
\textbf{\system Local:}
This app can be implemented as an OS core service, or a regular app.
The former implementation requires changes in Android design, and thus impacts all stakeholders involved in the chain, making it more difficult to roll-out.
However, device and OS manufacturers 
could opt for such choice and include \system Local in their customized OS.
In the latter implementation, the app can be standalone, or included in other apps (\eg,~a phone provider main app).

The limitation of using the regular app option is the \emph{App Standby Bucket} system introduced by Android 9 to manage app's requests for resources depending on their frequency of usage.
Apps that are not essential (\ie,~whitelisted by the phone vendor or OS) and/or have not been used in a while by the user receive lower priority or placed on standby.
As Local does not require user interaction, the app would receive lower priority.
Therefore, the app may not be immediately responsive to system notifications, and may not execute, or significantly delay tasks and message processing. %
This issue can be solved with two approaches:
1) Local is whitelisted in the unrestricted bucket list, \ie,~the common case of apps installed by phone vendors.
2) Local is included in a frequently used app to remain in the high priority bucket.
Regardless of this decision, the system design of \textsection~\ref{sec:design} remains same.

Currently, we implement Local as a standalone app and use the whitelisting approach.
The implementation counts 2,414 lines of Java code and uses the following:
\begin{itemize}[leftmargin=10pt,noitemsep,topsep=0pt]
    \item \system Library v0.1.0, to handle app authentication, inter-app communication, on-device ML training.
    \item Retrofit v2.9.0, an HTTP networking library for managing the communication with the back-end's REST API.
    \item Pushwoosh Android SDK v6.3.5 as a Push Notification library for receiving and parsing the push notification messages sent by the back-end.
\end{itemize}

\noindent
\textbf{Inter-App Communication:}
This functionality is required to exchange messages (\ie,~dictionary-based numerical and text) and data (\ie,~in binary format such as training samples and model parameters) between \system-enabled apps and Local.
In Android, inter-app communication for both messages and data can be achieved with explicit, manifest-declared \emph{Broadcast Receivers}. 
Similar communication can also be achieved with the \emph{ContentProvider API} or Android 10's only \emph{Scoped Storage}.
In our implementation, we use the approach of Broadcast Receivers for simplicity.

\noindent
\textbf{App Workers:}
For executing ML tasks in the background, we utilize Android's WorkManager API, a robust framework for executing tasks (workers) asynchronously for optimal usage of device resources.
All workers are executed sequentially when they belong to the same app.
The only exception is the on-device training per app in JM mode, where the OS can decide to execute in parallel.
Delays between workers can be introduced by the OS for better battery performance.
We identify these delays or pauses in our experiments with real devices, discussed next.

%% file: sections/06_evaluation.tex
\section{Experimental Evaluation}
\label{sec:experiments}

This experimental evaluation aims to answer the following questions:
a) How does \system operate in the wild, on real user Android devices?
b) Is it easily deployable? And does it function well with respect to device availability?
c) What do users think of the system's impact on their devices?
d) What is the \system ML performance and system overhead while computing FL projects on real user devices?
e) How does this performance compare with ideal conditions assumed in past studies with FL simulations?
f) How does the cross-app, on-device FL training perform?

\subsection{Experimental Setup}
\label{sec:exp-setup}

\subsubsection{Third-party apps}

We create three simple apps (named Red, Green and Blue, or $RGB$), with 101 lines of Java code each.
Each app includes \system Library v0.1.0 for supporting system functions and to demonstrate the simplicity and efficiency of a third-party app executing \system projects.

\subsubsection{Dataset}
\label{sec:dataset}

As a dataset for the ML model training and testing, we use $CIFAR$-$10$~\cite{cifar10}, a benchmark dataset commonly employed by FL researchers.
We do not anticipate drastic changes in system overhead if we use other datasets.
This dataset consists of $60k$ total samples of images ($32$$\times$$32$$\times$$3$) representing one of 10 types of objects, or classes.
We partition the training set into subsets of 150 samples, and each subset is assigned to an instance of a third-party app (Red, Green and Blue) to be installed on a user device.

Non-Independent and Identically Distributed (NonIID) data provide a more realistic view of how data would be distributed 
a) across users: real users cannot be expected to have examples of all classes of data;
b) across apps: different apps would have different classes that either overlap or not, and this can be persistent across users.
Thus, 4 different data distributions are constructed from $CIFAR$-$10$ data:
\begin{enumerate}[noitemsep,topsep=0pt,leftmargin=0.5cm]
    \item Independent and Identically Distributed (\emph{IID}): All apps have samples of all 10 classes in the data;
    \item NonIID Per User (\emph{NonIIDPU}): Each user has samples from 3 random classes, consistently across its apps;
    \item NonIID Per App Overlapping (\emph{NonIIDPAO}): Each app has 3 random classes that can overlap between the apps.
    \item NonIID Per App Persistent (\emph{NonIIDPAP}): 10 classes are split across the 3 apps: R \& G get 3 classes, B gets 4 classes; the class selection is persistent across all users.
\end{enumerate}

\subsubsection{ML model \& hyperparameters}

As a base DNN model we use MobileNetV2~\cite{s2018mobilenetv2}, which is pre-trained with ImageNet~\cite{imagenet_cvpr09} dataset with image size of 224x224.
As a head of the DNN (used for model personalization on user local data), we use a single dense layer with 32 neurons and $RELU$ activation function, followed by $softmax$ activation function (SGD optimizer with 0.001 learning rate, 0.0001 kernel regularizer, and 0.0001 for bias regularizer).
Furthermore, we applied typical parameter values used in other $FL$ works with $CIFAR$-$10$: 20 samples per batch and 20 epochs per batch.
We experimented with 10 up to 30 FL rounds.

\subsubsection{Performance Metrics}

We measure model performance using Accuracy when testing the global FL model on 1,000 unseen samples of $CIFAR$-$10$.
This ML performance is 1)~measured on FL global models built in ideal scenarios (via simulation) when all user devices participate in all FL rounds, and 2)~compared to real scenarios with real user devices freely joining and completing or dropping their FL rounds depending on their actual phone usage.
The simulation environment uses Keras TensorFlow v2.6.0 under Python v3.8.12.

We assess the cost of on-device \system functionalities by measuring computing cost via CPU utilization (in \%), time delay (in minutes), and battery discharge (in mAh) taken to compute said ML tasks.
These metrics are computed in a course-grained fashion on real user devices while training ML models, as well as in a fine-grained, lab-controlled fashion on test devices using the BatteryLab infrastructure~\cite{batterylab} that utilizes \emph{Monsoon High Voltage Power Monitor}~\cite{msoon} for accurately measuring power discharge of a phone's battery.

\subsection{How does \system operate in the wild?}
\label{sec:exp-user-study}

We assess system deployability and overhead by designing a user study and deploying the various \system components on real user devices.
Recruited users are paid for their participation over two weeks providing their consent for:
\begin{enumerate}[leftmargin=14pt,noitemsep,topsep=0pt]
    \item Installing 4 mobile apps: \system Local and RGB apps, each preloaded with 150 $CIFAR$-$10$ samples for each of the 4 data distributions.
    \item The \system Server to receive regular status from their device, to decide if it is available for local ML training inside the third-party apps and \system Local.
    \item The \system Server to invite their device to join FL projects and perform ML training on the preloaded data.
\end{enumerate}
Note: the ML models are never trained on user personal data; only on $CIFAR$-$10$ sets of pre-allocated samples in each app installed by users.
In fact, this user study followed strict principles of ethical information research and use of shared measurement data~\cite{dittrich2012menloreport,rivers2014ethicalresearchstandards}, as well as our institution's ethical, privacy and security policies, in compliance with GDPR on getting informed consents from users for handling their data.

\subsubsection{What is the device availability for \system tasks?}

We actively recruit users for two weeks, asking each to participate for at least 2 weeks to allow us enough time to perform essential ML training tasks.
However, users are free to join and exit the study at any moment creating a dynamic set of client devices available for \system-related ML tasks.
The overall duration of the user study is 4 weeks, with a total of 144 unique users from 5 EU countries sharing the same timezone, joining and installing the \system apps for at least 1 day.
We observe a fairly balanced breakdown of users with respect to gender (61\% male, 39\% female) and age groups (18-29y: 34\%, 30-39y: 41\%, 40-49y: 17\%, 50+y: 8\%).
Furthermore, we identify 116 unique models of Android phones, from 15 unique smartphone manufacturers.
The top 3 are Samsung (44\%), Xiaomi (24\%) and Huawei (12\%).

\begin{figure}[t]
    \centering
    \begin{subfigure}[b]{0.49\textwidth}
        \centering
        \includegraphics[width=\textwidth]{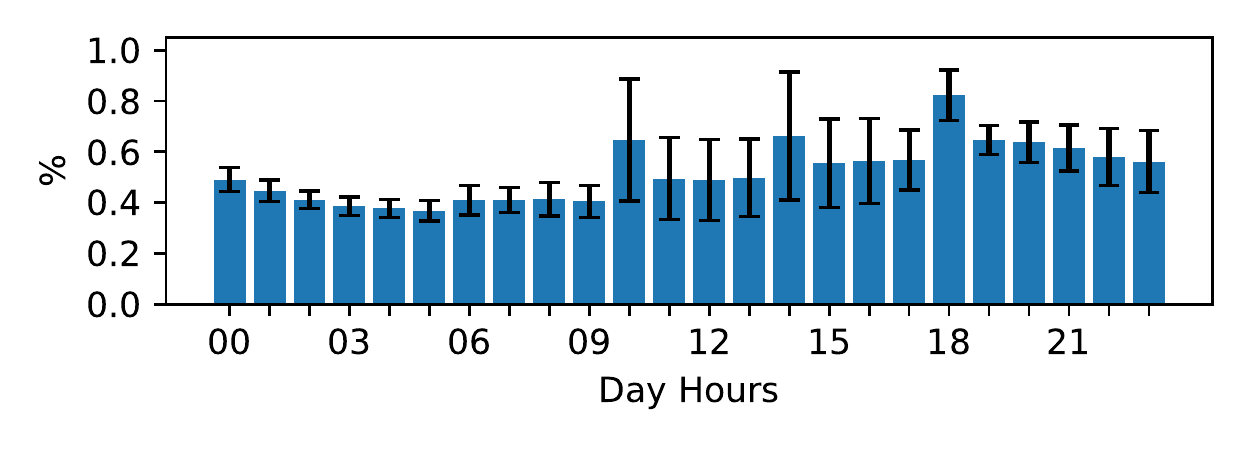}
        \label{fig:responses-per-hour}
    \end{subfigure}
    \vspace{-1cm}
    \caption{Portion of unique devices available for FL training per hour, during the study.
    Error bars: 1 st. deviation.
    }
    \label{fig:user-study-responses}
\vspace{-0.4cm}
\end{figure}

\begin{figure*}[ht]
    \centering
    \begin{subfigure}[b]{0.24\linewidth}
        \centering
        \includegraphics[width=\linewidth]{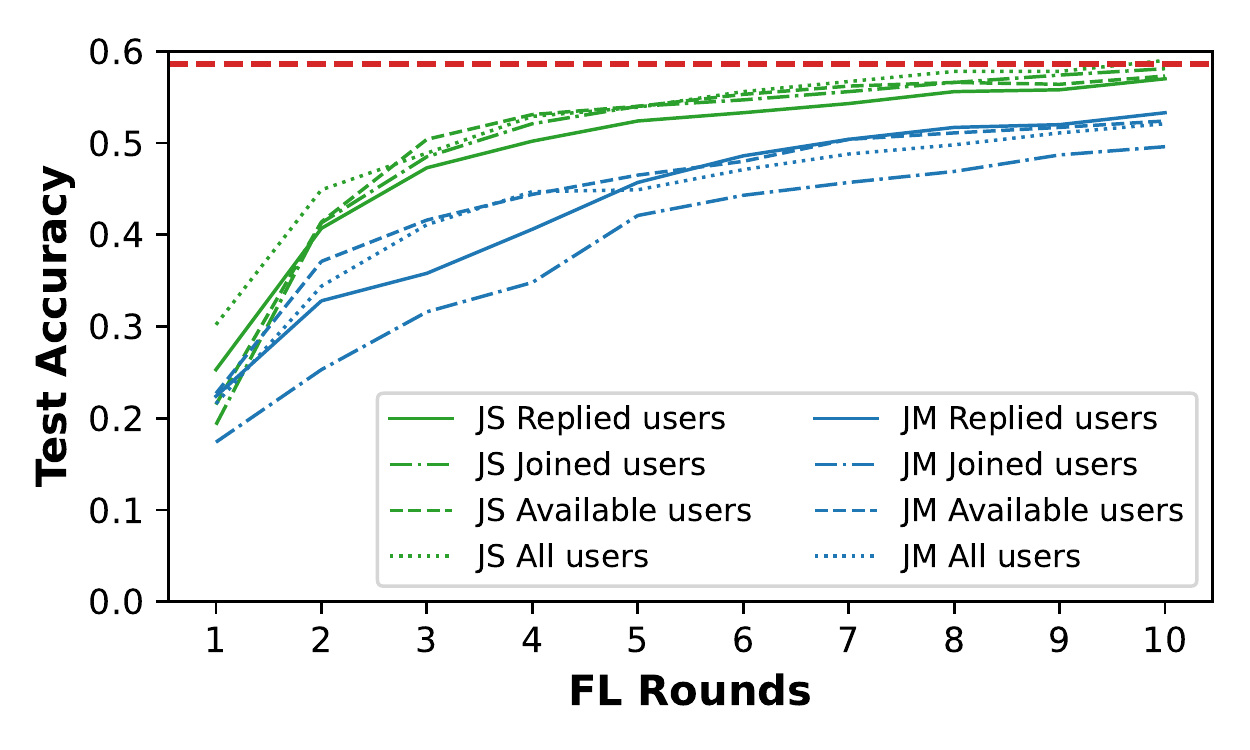}
        \caption{IID}
        \label{fig:ml-performance-iid}
    \end{subfigure}
    \hfill
    \begin{subfigure}[b]{0.24\linewidth}
        \centering
        \includegraphics[width=\linewidth]{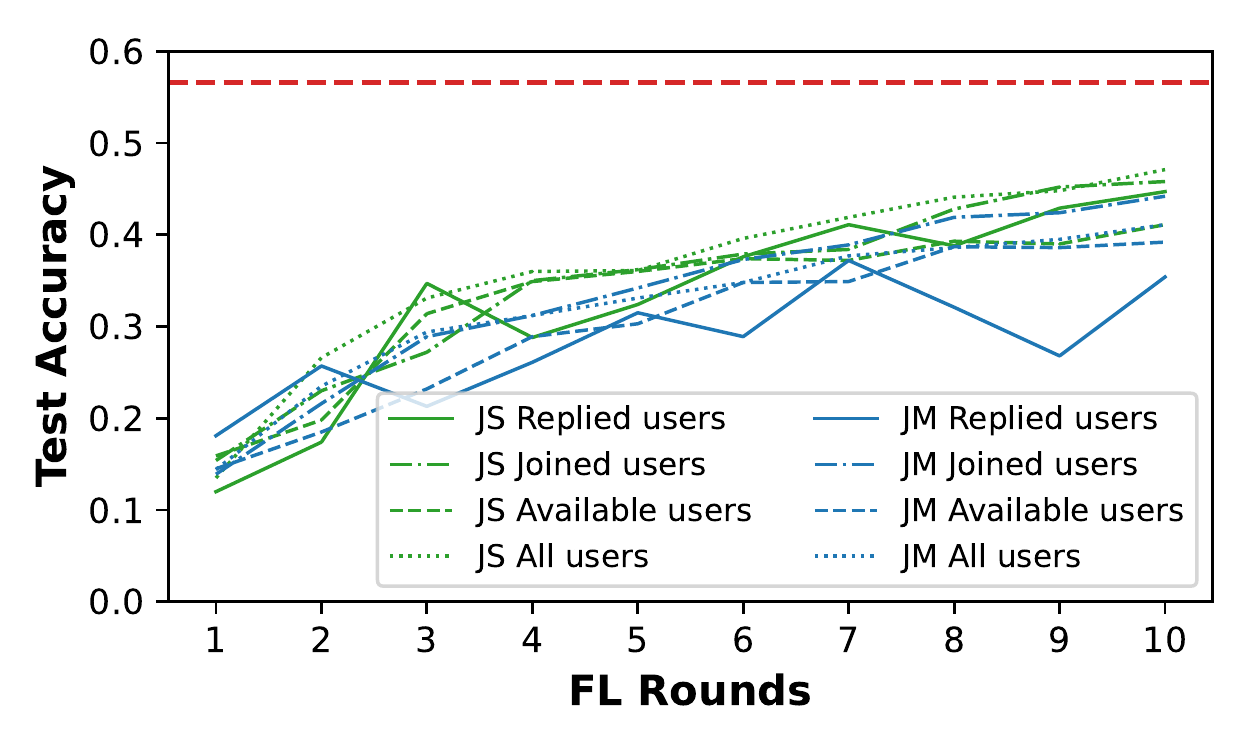}
        \caption{NonIIDPU}
        \label{fig:ml-performance-noniid-peruser}
    \end{subfigure}
    \hfill
    \begin{subfigure}[b]{0.24\linewidth}
        \centering
        \includegraphics[width=\linewidth]{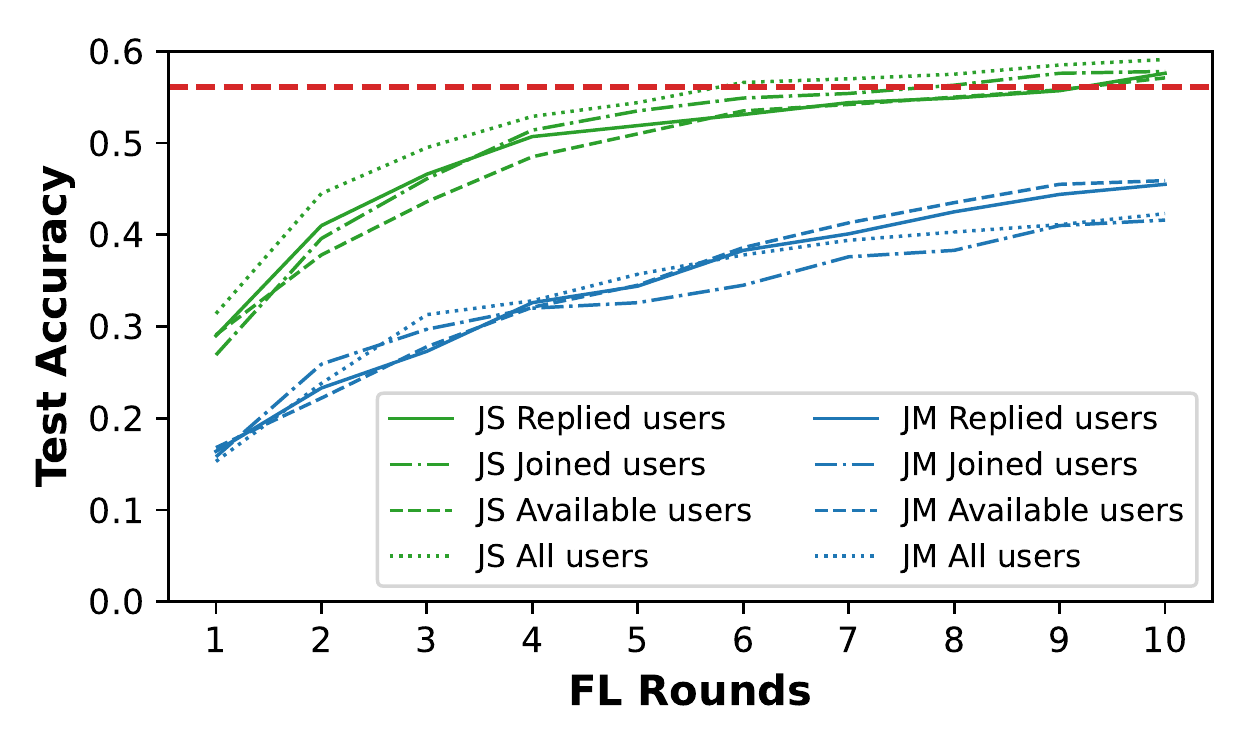}
        \caption{NonIIDPAO}
        \label{fig:ml-performance-noniid-perapp-overlapping}
    \end{subfigure}
    \hfill
    \begin{subfigure}[b]{0.24\linewidth}
        \centering
        \includegraphics[width=\linewidth]{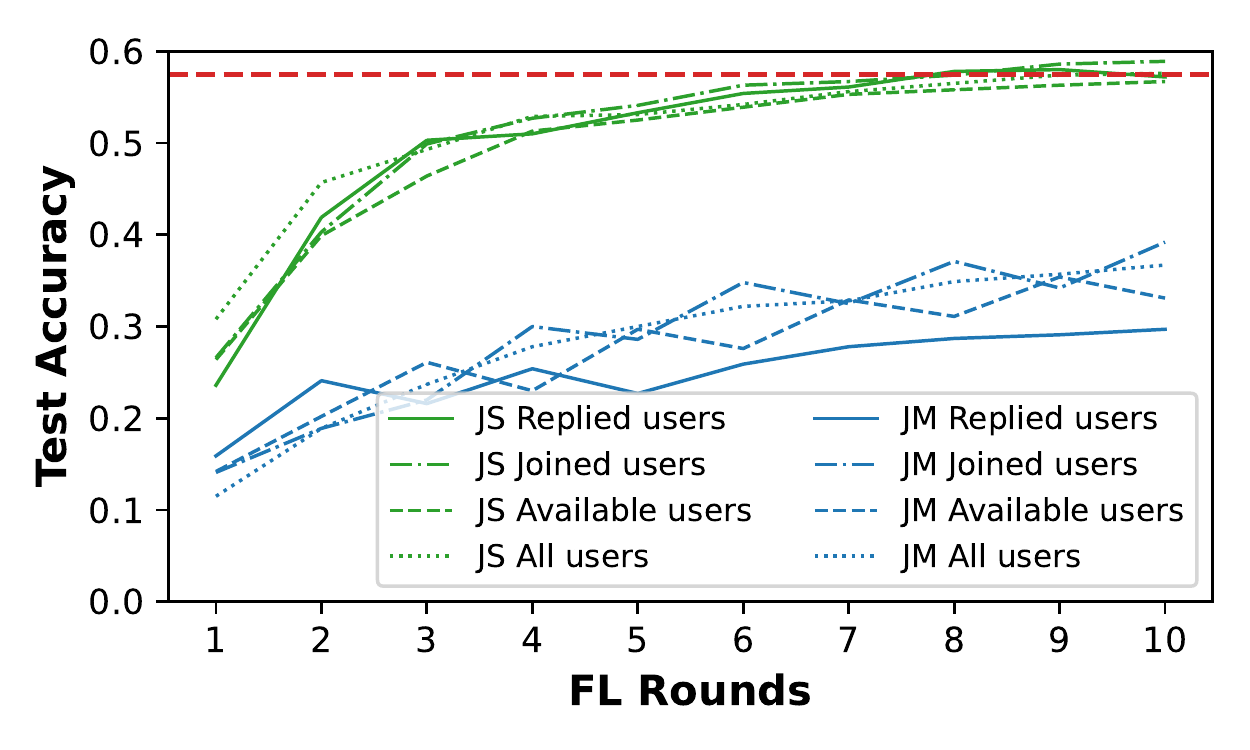}
        \caption{NonIIDPAP}
        \label{fig:ml-performance-noniid-perapp-persistent}
    \end{subfigure}
    \hfill
    \vspace{-0.3cm}
    \caption{Test accuracy per FL round, for different experimental configurations: IID, NonIIDPU, NonIIDPAO, NonIIDPAP, with Joint Samples (JS) vs. Joint Models (JM).
    Horizontal dashed line: Baseline with centralized ML on all data.}
    \label{fig:user-study-ml-performance}
\end{figure*}

We found an average of 94-99\% of total (or 76-94 unique) devices available per day for \system ML tasks.
This ratio was fairly constant across all days of the week.
When studying device availability on a more granular level (\ie,~per hour, Fig.~\ref{fig:user-study-responses}), we find that there are 37-82\% of total (or 33-73 unique) devices available for training at any given hour of the day.
We observe a slight diurnal behavior of user devices (since they are a proxy of users being active on their phones), with a slight dip in availability during early morning hours, and an increase during work and dinner hours.

\subsubsection{What is the user perception of \system impact on devices?}
Users are also prompted 3 times per day to fill-in a simple user survey of 4 questions to provide feedback on (i) their mobility through the day (scale 1-5), (ii) their phone usage through the day (scale 1-5), (iii) their phone performance through the day (with 3 possible answers, \eg,~if during the day, their phone performed slower or was heating more often than usual) and (iv) the battery performance through the day (with 3 possible answers, \eg,~if during the day, their phone was running out of battery more often than usual).

We observe these prompts produced 3 corresponding spikes in device availability at 10am, 2pm and 6pm CET (Fig.~\ref{fig:user-study-responses}).
We collect $3780$ survey responses across users and time.
In summary, $90.8\%$ remarked no change in phone performance, $8.5\%$ noticed some change (device was slower/warmed-up a bit), and only $0.8\%$ noticed a drastic change (device was very slow/warmed-up a lot).
Also, $73.7\%$ remarked no change of phone battery, $22.7\%$ noticed their battery was somewhat lower than usual and only $3.6\%$ noticed a drastic change on battery/had to charge it more than usual.
Perhaps expectedly, we noticed a significant correlation between users' declared mobility and usage of phone ($\rho$=0.365, p-value$<$$3e^{-16}$), stating the obvious: users use their phone more often when they moved.
Similarly, we find a significant correlation between users' declared phone overall performance and phone battery consumption ($\rho$=0.413, p-value$<$$2.2e^{-16}$).
These results confirm that users who noticed their phone having performance slowdown, also found their battery performing worse.

\noindent
\textbf{Summary:}
Overall, we find more than 94\% of registered devices available per day for ML training tasks ($\sim$33-73 unique devices at any hour of day), and a slight diurnal behavior of user devices in their availability.
Also, users appear to be OK with the way their phone is performing during the day and while \system ML training tasks are executed at the same time as their daily phone tasks.
These results demonstrate that running FL model training on real users' phones is practical when utilizing available OS APIs, even with the introduced restrictions that affect the duration of a FL round.

\subsection{How well can FL models train in the wild?}
\label{sec:exp-ml-perf}

The user study aims to investigate to what extend real devices can execute FL tasks with a variable number of devices.
During the study, we execute more than 50 different FL projects, while varying experimental parameters such as:
\begin{itemize}[leftmargin=10pt,noitemsep,topsep=0pt]
    \item type of local ML modeling: single-app vs. cross-app
    \item type of cross-app modeling: $JS$ vs. $JM$
    \item number of apps participating in training: \{1, 2, 3\}
    \item type of data/app: \mbox{IID}, \mbox{NonIIDPU}, \mbox{NonIIDPAO}, \mbox{NonIIDPAP}
    \item FL round duration: \{1, 1.5, 2, 8, 12, or 24\} hours
\end{itemize}

Due to limited space, we only present results from the more challenging scenarios, involving cross-app FL modeling with 3 apps (RGB) computing models, in JS or JM mode, using IID, NonIIDPU, NonIIDPAO, NonIIDPAP data. %

\begin{figure*}[ht]
    \centering
    \begin{subfigure}[b]{0.57\linewidth}
        \centering
        \includegraphics[width=\linewidth]{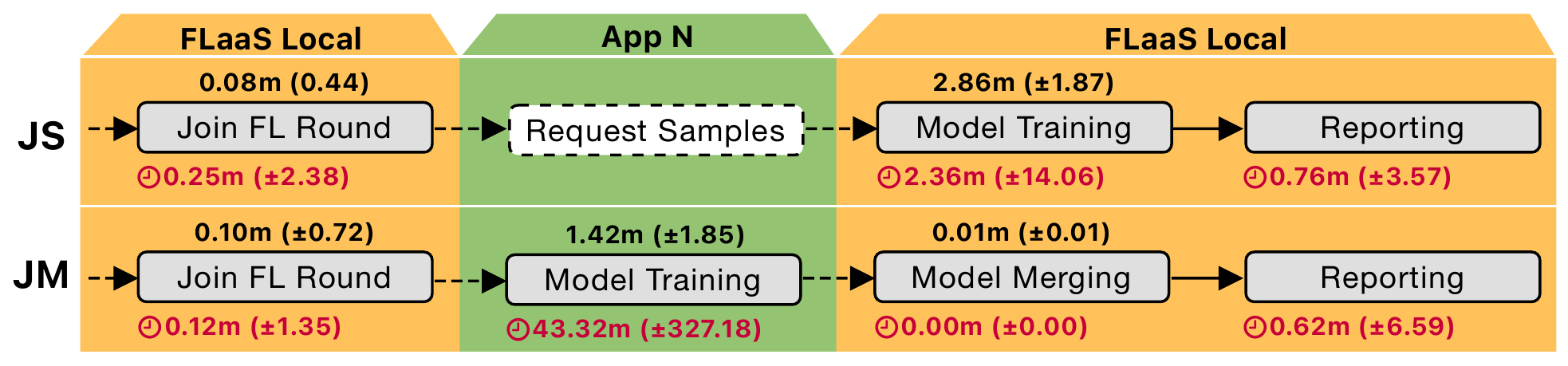}
        \caption{Available app workers, with average time (and Std.) for task duration (black) and task waiting time (red).}
        \label{fig:workers}
    \end{subfigure}
    \hfill
    \begin{subfigure}[b]{0.21\linewidth}
        \centering
        \includegraphics[width=\linewidth]{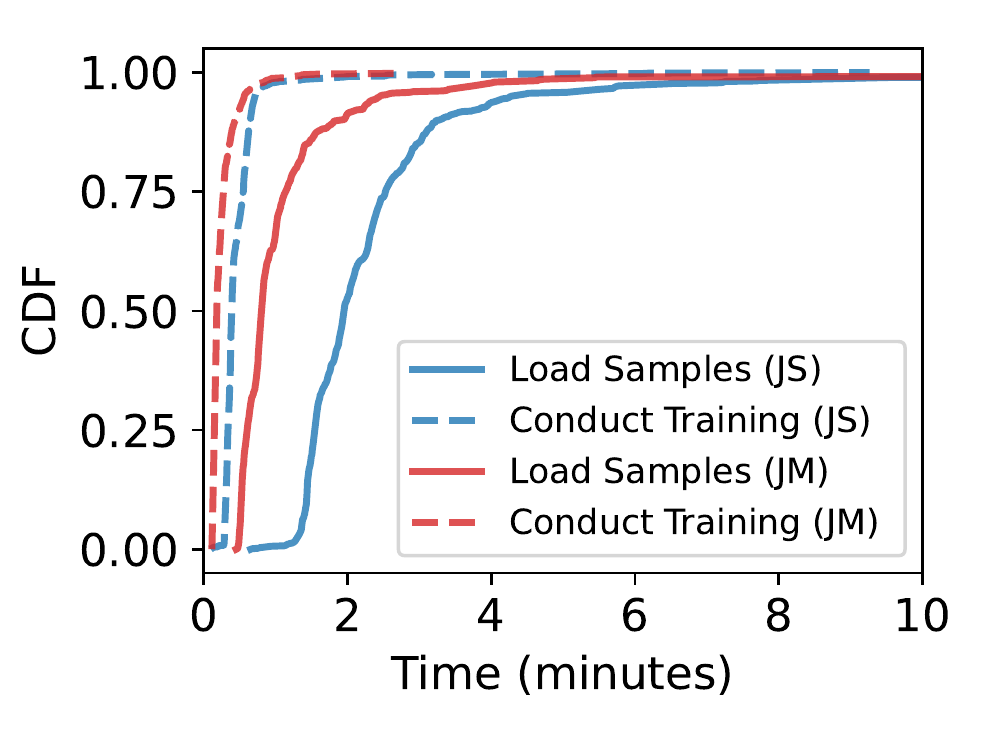}
        \caption{Model Training}
        \label{fig:user-study-cdfs-training-task}
    \end{subfigure}
    \hfill
    \begin{subfigure}[b]{0.21\linewidth}
        \centering
        \includegraphics[width=\linewidth]{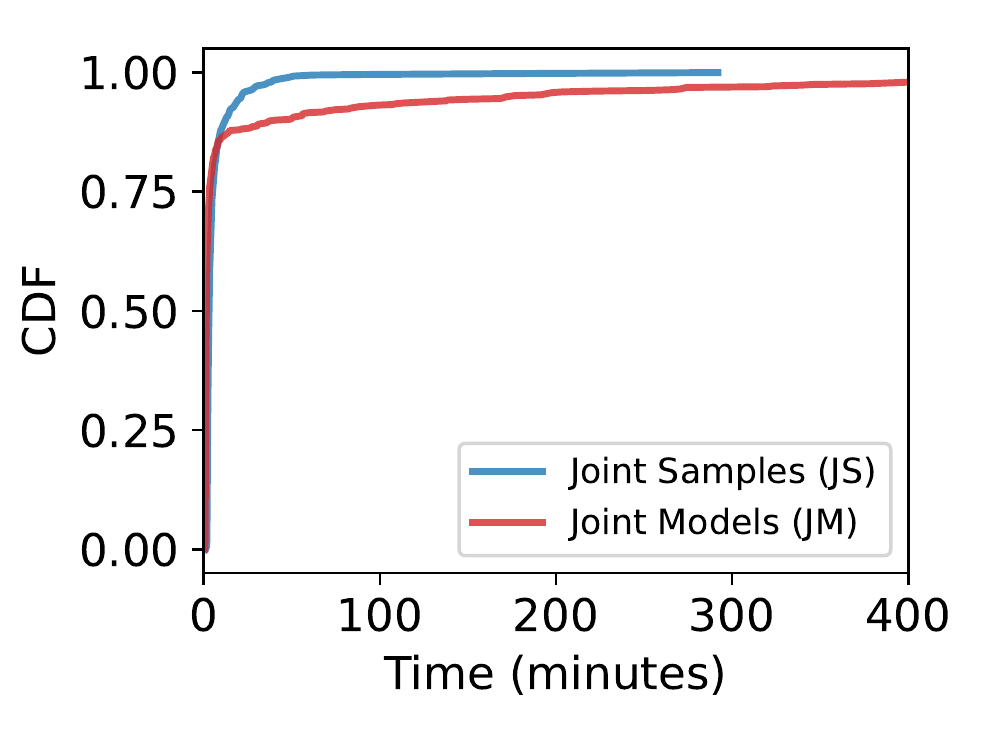}
        \caption{Total Duration}
        \label{fig:user-study-cdfs-total}
    \end{subfigure}
    \caption{Duration of on-device functions of devices participating in FL round using JS or JM mode.
    (a) Available app workers (grey boxes) execute different tasks with duration declared with black numbers; worker dependency can be chained (black arrows) and has task execution waiting time declared with red numbers.
    Inter-app communication (dashed arrows) is instantaneous, but with restrictions: OS does not allow a process to take many resources for too long.
    (b) CDF of the most expensive worker (\ie, Model Training) for the JS or JM modes, decomposed in two main tasks: Load Samples and Conduct Training.
    (c) CDF of total duration of on-device pipeline, for the two modes.
    }
    \label{fig:user-study-cdfs-js-jm-totalduration}
\end{figure*}

\subsubsection{How does user availability impact model performance?}
We compare test accuracy achieved per configuration (under 90min FL rounds) on the \system infrastructure using incrementally smaller subsets of user devices:
\begin{itemize}[leftmargin=10pt,noitemsep,topsep=0pt]
    \item All registered user devices ($\sim$100 at any moment) are assumed available and join FL project rounds (results from simulation on GPU cluster).
    \item Only user devices that reported availability are assumed to join FL project rounds; on average 30-70 devices per hour (results from simulation on GPU cluster).
    \item Only user devices that actually joined an FL project, but not necessarily reported a model; on average, 30-70 devices per hour (results from simulation on GPU cluster).
    \item Only user devices that were available, joined the FL project and actually reported a model trained for the FL project; $\sim$20-40 devices per hour (results from real user devices).
\end{itemize}

Figure~\ref{fig:user-study-ml-performance} shows that across all scenarios, with a few tens of devices joining an FL project, \system can produce a useful model and comparable to a centralized ML model.
In particular, even though the number of user devices joining and delivering ML models is much lower than the ideal, each model trained performs similarly.

\subsubsection{How does data distribution impact model performance?}

We note that in JS mode, and expectedly, IID (Fig.~\ref{fig:ml-performance-iid}) leads to ML models with $21.58\%$ higher accuracy on Replied users, than with NonIIDPU (Fig.~\ref{fig:ml-performance-noniid-peruser}), at the $10^{th}$ round.
NonIIDPAO (Fig.~\ref{fig:ml-performance-noniid-perapp-overlapping}) and NonIIDPAP (Fig.~\ref{fig:ml-performance-noniid-perapp-persistent}) also perform well; in some cases slightly better ($1.05\%$ and $0.35\%$, respectively) than IID.
Results in JM mode follow a slightly different pattern. IID performs best, NonIIDPAO performs worst (\ie,~$14.63\%$ lower than IID), and NonIIDPAP $44.28\%$ lower compared to IID.
Interestingly, the performance of NonIIDPU under JM is rather unstable across rounds, with best performance at round 7 (\ie.~$31.21\%$ lower than IID).
With further investigation, we identified that this behaviour depends on the number of epochs executed locally per round.

\subsubsection{How does cross-app model merging perform?}

Figure~\ref{fig:user-study-ml-performance} shows a higher ML performance of models trained by sharing samples with \system Local (\ie,~$JS$), vs. training individual models, one per third-party app, and then averaging them at Local (\ie,~$JM$).
In fact, $JS$ mode at the 10$^{th}$ round leads to a model with $6.49\%$ higher test accuracy than $JM$ mode for IID, $20.81\%$ higher for NonIIDPU, $21.01\%$ higher for NonIIDPAO and $48.07\%$ higher for NonIIDPPA.
The decreased performance of NonIIDPAP is expected since it enforces the distribution of data across apps, while NonIIDPAO allows the cross-silo effect to be less prominent.
Additionally, the improved performance of JS is expected since sharing data with Local allows it to train the joint model on $3\times$ more data and thus optimize it better, than averaging models from the RGB apps, trained on 1/3 of samples.

The overall JS performance is comparable with our baseline non-FL approach where all data are shared in a centralized ML service for training (early stopping when monitoring validation loss with validation\_split=0.3, patience=5 and min\_delta=0.001), with only exception the more challenging NonIIDPU that performed $26.62\%$ lower. The decreased performance of JM compared to baseline is more obvious: $6.19\%$ for IID, $59.89\%$ for NonIIDPU, $23.30\%$ for NonIIDPAO and $93.60\%$ for NonIIDPAP.

\noindent
\textbf{Summary:}
The ML performance results show the feasibility of our design. With a few 10s of devices reporting trained models, \system builds a useful FL model, comparable to ideal scenarios of all devices participating and reporting models, and even centralized ML training.
We also show that cross-app modeling with shared samples, in comparison to joint models, achieves, on average, a $24.10\%$ higher test accuracy across all data distributions.
Expectedly, NonIID data across users (NonIIDPU), while being a more realistic scenario, leads to models with lower accuracy.

\subsection{What is the cost of \system operations?}
\label{sec:exp-device-cost}

\subsubsection{Latency of \system functions on user devices?}

We showed that real user devices with their intermittent participation in FL rounds can enable \system to build a well-performing model.
Next, we assess on-device cost for 3 fundamental \system functions executed during an FL round, by measuring time required to execute them on user devices:
\begin{itemize}[leftmargin=12pt,noitemsep,topsep=0pt]
    \item \emph{Join FL round:} Communicate with Server and download global FL model parameters (Fig.~\ref{fig:workers}).
    \item \emph{Model Training:} Load local samples into on-device ML engine and conduct training (Fig.~\ref{fig:workers} and Fig.~\ref{fig:user-study-cdfs-training-task}).
    \item \emph{Model Merging:} Aggregate local models and apply FedAvg (in JM only; typically very small. Fig.~\ref{fig:workers}).
    \item \emph{Reporting:} Communicate with Server to report updated local model and other performance metrics (Fig.~\ref{fig:workers}).
\end{itemize}

Figure~\ref{fig:workers} presents the combination of app workers and in-app communication pipeline for executing FL tasks under JS or JM training modes.
The task begins when a notification is received by the device that immediately schedules the first app worker (\ie,~Join FL Round) and continues with an inter-app communication with the participating apps.
As expected, the most expensive task is Model Training under both JM and JS.
Interestingly, though JS takes $2$x more time than JM for this step, the waiting time for the task to be executed in JM is $18.4$x slower.
This is because the training is to be executed on three toy apps (RGB), which remain mostly inactive (\ie,~no user interaction takes place).
Thus, the OS, while optimizing task execution for power efficiency, schedules such long tasks later.
However, we expect this time to be significantly lower when apps are actively used.
For all other workers, both task duration and waiting time are short. 

In Figure~\ref{fig:user-study-cdfs-training-task}, we breakdown the execution of Model Training and measure average time needed to load samples and conduct model training.
Note: results are independent of data distribution (IID, NonIIDPU, NonIIDPAO, NonIIDPAP), but dependent on type of cross-app modeling ($JS$ vs. $JM$).
Interestingly, the most costly operation is to load data into the on-device training engine, with median time $1.94$ and $0.80$ minutes, for $JS$ and $JM$, respectively.
Finally, there is the expected difference when conducting ML training for $JS$ and $JM$ (median $0.39$ and $0.18$ minutes, respectively).

Finally, in Figure~\ref{fig:user-study-cdfs-total}, we study the total time to complete an FL round, for the cross-app modeling ($JS$ vs. $JM$).
This time includes potential pauses of the FL training process, if, for example the device OS delays the worker execution (red numbers Fig.~\ref{fig:workers}).
Interestingly, the great majority of FL rounds, in both $JS$ and $JM$ have short duration: the 80$^{th}$ percentile duration of an FL round is $6.15$ minutes for $JS$ and $4.65$ minutes for $JM$.
This means \system or other FL-based ML systems can train full FL rounds in short periods of time.
However, straggler devices needing 100s of minutes to complete their FL tasks, create a long tail: we find that 90$^{th}$(99$^{th}$) percentile for the two modes is $12.1$($48.7$) minutes for $JS$ and $40.3$($2527.3$) minutes for $JM$.
These results motivate novel Asynchronous FL systems~\cite{huba2022papaya-mlsys}.

\begin{table}[t]
\centering
\caption{Device Specification (CPU and Memory)}
\label{tab:specification-devices}
\footnotesize
\begin{tabular}{ ll }
 Model & Device Specification \\
  \midrule
 Pixel~3a (P3a) & 8-core (2x2.0 \& 6x1.7 GHz), 4GB RAM \\
 Pixel~4 (P4)  & 8-core (1x2.84, 3x2.42 \& 4x1.78 GHz), 4GB \\
 Pixel~5 (P5)  & 8-core (1x2.4, 1x2.2 \& 6x1.8 GHz), 8GB RAM \\
\end{tabular}
\end{table}

\subsubsection{Device cost of \system functions?}

Diving deeper into the system cost of on-device ML training, we perform several experiments with three popular Google Pixel devices (specs in Table~\ref{tab:specification-devices}) as test devices in a controlled setting: the devices are forced to perform the ML task on IID data, from beginning (received notification of new ML task) until the end (delivery of ML model to Server) without any breaks.
All devices were updated to Android 12 and with deactivated OS-related automated process that would influence measurements (\ie,~automatic updates, adaptive battery and brightness).
During evaluation, all devices are connected to a stable 5GHz (IEEE 802.11ac) Wi-Fi network, while CPU utilization was collected using the Android Device Bridge (ADB) tool, and Battery Discharge using the BatteryLab~\cite{batterylab} infrastructure.
All RGB apps are whitelisted (\ie,~set to Unrestricted Battery usage) aiming to measure the most optimal scenario where all apps are responsive to ML training requests, without further scheduling delays introduced by Android's Work Manager.

Table~\ref{tab:cost-devices} shows the average time taken for a device to perform the ML training task, for the three test devices and two cross-app modes ($JS$ vs. $JM$) with a total of 450 samples.
As expected, the older device (Pixel 3a) takes the longest to finish, regardless of modeling mode.
Also, confirming results from the user study, $JS$ is slower than $JM$ since apps have to first transfer data to Local, and then Local has to train the model on them.
On the other hand, $JM$ builds 3 individual models, even in parallel if the OS's Work Manager allows it.
Interestingly, this parallelized execution in $JM$ is reflected in 
CPU usage, which is, on average, $74.7\%$ higher on $JM$ than $JS$, across devices.
Finally, %
the reduced execution time, even at higher CPU usage, leads to overall reduction in consumed power from $JM$: $41\%$ lower than $JS$.

\begin{table}[t]
\centering
\caption{Cost for on-device ML training: Training duration, CPU usage, and Power discharge.
Means and standard deviation (SD) computed on distribution of each metric on each device and modeling mode, over 10 rounds.}
\vspace{-0.3cm}
\label{tab:cost-devices}
\footnotesize
\begin{tabular}{
            p{0.2cm}
            p{1.0cm}p{0.9cm}
            p{0.9cm}p{0.9cm}
            p{0.9cm}p{0.9cm}}
                                            &
\multicolumn{2}{c}{Duration(sec)}     &
\multicolumn{2}{c}{CPU usage(\%)}     &
\multicolumn{2}{c}{Discharge(mAh)}   \\
\midrule
            &   \multicolumn{1}{c}{JS}  &
                \multicolumn{1}{c}{JM}  &
                \multicolumn{1}{c}{JS}  &
                \multicolumn{1}{c}{JM}  &
                \multicolumn{1}{c}{JS}  &
                \multicolumn{1}{c}{JM}  \\
            &   Mean(SD)  &   Mean(SD)  &
                Mean(SD)  &   Mean(SD)  &
                Mean(SD)  &   Mean(SD)  \\
\midrule
P3a    &   167.3(13.1)  &   85.3(1.5)    &
                17.9(2.1)    &   29.7(0.6)    &
                56.8(8.6)   &   37.8(0.5)     \\
P4     &   117.1(1.1)   &   62.1(0.8)    &
                15.4(0.1)    &   27.8(0.4)    &
                41.3(0.9)   &   29.7(0.5)     \\
P5     &   115.0(0.4)   &   62.8(0.9)    &
                16.1(0.1)    &   28.8(0.4)    &
                42.7(1.0)   &   32.4(0.7)     \\
\end{tabular}
\end{table}

\noindent
\textbf{Summary:}
Overall, a regular user device is expected to spend small amount of time for completing an FL round: a median of $131.7$s and $69.6$s in optimal case (test devices) and $178.2$s and $85.2$s in regular case (real users), for $JS$ and $JM$, respectively.
Notably, there are straggler devices requiring significantly more time; a practical FL system should stop waiting for them very early on in the FL round.
Also, the most costly function in the process is loading data for the ML training.
Future work could further optimize this step.
Further, $JS$, in comparison to $JM$, requires $90.00\%$ more execution time, but uses $74.68\%$ less CPU.
However, it requires $40.96\%$ more power to process the same number of samples but in a single app (\system Local).
Finally, older devices (\eg,~Pixel 3a) require more execution time ($44.2\%$ and $36.6\%$ increase) and power ($35.2\%$ and $21.7\%$ increase) to perform the same ML task in JS and JM, respectively.

It is worth mentioning that in the Android ecosystem, and in contrast to Apple's iOS, on-device training is only available through TensorFlow Lite with limitations: only utilizing CPU (and not available GPU) and no support of multi-processing.
Future releases from Google TensorFlow could allow great performance increases, as indicated in their roadmap~\cite{tflite-roadmap2021}.

%% file: sections/02_background.tex
\section{Related Work}
\label{sec:background}

The industrial, research and open source communities have proposed various frameworks for distributed or federated learning.
On the industrial side, large Internet companies, such as Google, Meta and Apple, are using FL on Android and iOS phones for different tasks such as keyboard next word prediction, error correction, and acoustic keyword trigger~\cite{bonawitz2019FL-sysml,huba2022papaya-mlsys,apple-fl}.
Interestingly, Papaya from Meta~AI~\cite{huba2022papaya-mlsys} performs buffered asynchronous FL for next-word prediction using the concept of staleness to incorporate late arrivals of models from devices into the global FL model.
Overall, these frameworks are tailored for mobile devices, but they are designed on a per-app basis and embedded in company products.

Training model with FL across companies (\ie,~cross-silo FL) has also received attention from major industry players, such as Nvidia, IBM, and Intel, as well as startups with focus on healthcare or finance~\cite{bonawitz2021,sherpa-ai}. 
While these frameworks are relevant to our work, they are proprietary and not designed for generalized use in mobile apps.
Similarly, other startups provide proprietary frameworks for FL or distributed training, but not for mobile environments~\cite{integrate-ai}.
Finally, all these efforts, including the cross-silo FL solutions, do not tackle the cross-app problem of how to enable third-party apps to collaborate and jointly build an ML model that solves same or new ML problems, on the user device.
This paper proposes the first practical solution to enable this joint modeling between apps.

On the open-source community side, different frameworks have been proposed for decentralized, secured and private ML:
Open source frameworks developed for simulations, optimization, high level programming language, or benchmarking for FL and decentralized computation systems, such as TensorFlow Federated (TFF)~\cite{tensorflow-federated2020}, coMind~\cite{comind2019} or LEAF~\cite{leaf-benchmark2019}, though none of these frameworks have support for mobile environments.
Also, \textit{OpenMined}~\cite{openmined2020} proposes the libraries \textit{PySyft} and \textit{PyGrid} by employing $MPC$, homomorphic encryption, $DP$ and $FL$ for secured and $PPML$ modeling.
FedML~\cite{he2020fedml} proposes an FL product closer to \system, by allowing FL customers to seamlessly deploy their FL projects to edge devices and collaborate with other each other.

On academic side, Hermes~\cite{hermes-mobicom} and Flower~\cite{flower2021} are two frameworks close to this study's goals, allowing FL to be deployed on heterogeneous devices, such as phones and tablets.
However, Hermes focuses on providing personalization and communication efficiency of FL models while considering data heterogeneity.
Also, Flower supports Android-based devices using TFLite~\cite{tflite2022}, but assumes an gRPC connection always exists between server and FL clients, who are assumed to be always available for training.
This design decision makes it unsuitable for ad-hoc or asynchronous FL modeling on mobile devices we aim to support with \system.
Also, both Hermes and Flower do not support cross-app modeling, nor assume the \textit{as-a-service} paradigm, both which are essential design elements of \system.

\noindent
\textbf{Contributions:}
\system provides an easy to use, open-source framework for mobile app developers, currently not provided by past solutions.
Specifically, we advance the FL state-of-art by designing, implementing and evaluating:
i) a first of its kind middleware for mobiles enabling single/cross-app on-device training;
ii) a set of secure and PP mechanisms to enable cross-app training;
iii) a mobile app library and easy-to-use API for app developers to enable FL usage seamlessly;
iv) a robust end-to-end design for system scalability and in-the-wild deployment.
Furthermore, we are the first to provide an in-depth evaluation of the benefits of cross-silo FL in the context and at the level of mobile apps (not at cross-silo servers), and in-the-wild analysis of this proposal.%

%% file: sections/10_conclusions.tex
\section{Conclusion}
\label{sec:conclusion}

In this paper, we presented \system, the first practical Federated Learning system for mobile environments, that supports single and joint-app ML modeling, in a secure, private and ``as-a-Service'' fashion.
We implemented \system for Android-based mobile devices and performed extensive in-lab and in the wild experiments with 144 real users.
Our results show the feasibility of the \system design: with a few tens of devices, \system can build a useful ML model having an almost negligible impact on user phone experience and requiring a few minutes to perform each FL round.
We also showed the impact of different strategies for joint ML model training across third-party mobile apps and presented several insights on real devices with actual users' availability and opportunities for FL training.
The complete source-code of our system is available at: \url{https://github.com/FLaaSResearch}.

%% file: main.bbl
\begin{thebibliography}{10}

\bibitem{heroku}
Heroku cloud application platform.
\newblock \url{https://www.heroku.com}, 2021.

\bibitem{integrate-ai}
Machine learning for decentralized data.
\newblock \url{https://integrate.ai}, 2021.

\bibitem{pushwoosh}
Pushwoosh notification service.
\newblock \url{https://www.pushwoosh.com}, 2021.

\bibitem{sherpa-ai}
Sherpa.ai.
\newblock \url{https://www.sherpa.ai}, 2022.

\bibitem{aws-mlaas2020}
{\sc Amazon}.
\newblock Machine learning on aws.
\newblock \url{https://aws.amazon.com/machine-learning/}, 2020.

\bibitem{bagdasaryan2019ancile}
{\sc Bagdasaryan, E., Berlstein, G., Waterman, J., Birrell, E., Foster, N.,
  Schneider, F.~B., and Estrin, D.}
\newblock Ancile: Enhancing privacy for ubiquitous computing with use-based
  privacy.
\newblock In {\em Proceedings of the 18th ACM Workshop on Privacy in the
  Electronic Society\/} (2019), pp.~111--124.

\bibitem{bonawitz2019FL-sysml}
{\sc Bonawitz, K., Eichner, H., Grieskamp, W., Huba, D., Ingerman, A., Ivanov,
  V., Kiddon, C., Konečný, J., Mazzocchi, S., McMahan, H., Overveldt, T.,
  Petrou, D., Ramage, D., and Roselander, J.}
\newblock Towards federated learning at scale: System design.
\newblock In {\em 2nd SysML Conference\/} (2019).

\bibitem{bonawitz2021}
{\sc Bonawitz, K., Kairouz, P., McMahan, B., and Ramage, D.}
\newblock Federated learning and privacy: Building privacy-preserving systems
  for machine learning and data science on decentralized data.
\newblock {\em Queue 19}, 5 (oct 2021), 87–114.

\bibitem{leaf-benchmark2019}
{\sc Caldas, S., Meher Karthik~Duddu, S., Wu, P., Li, T., Konečný, J.,
  McMahan, H.~B., Smith, V., and Talwalkar, A.}
\newblock Leaf: A benchmark for federated settings.
\newblock \url{https://leaf.cmu.edu}, 2019.

\bibitem{comind2019}
{\sc CoMind}.
\newblock comind: Collaborative machine learning.
\newblock \url{https://comind.org}, 2019.

\bibitem{corinzia2021multi-taskFL}
{\sc Corinzia, L., Beuret, A., and Buhmann, J.~M.}
\newblock Variational federated multi-task learning.
\newblock {\em \url{https://arxiv.org/pdf/1906.06268.pdf}\/} (2019).

\bibitem{datafleets2020}
{\sc Datafleets}.
\newblock Datafleets: The federated intelligence platform.
\newblock \url{https://www.datafleets.com}, 2020.

\bibitem{imagenet_cvpr09}
{\sc Deng, J., Dong, W., Socher, R., Li, L.-J., Li, K., and Fei-Fei, L.}
\newblock {ImageNet: A Large-Scale Hierarchical Image Database}.
\newblock In {\em CVPR09\/} (2009).

\bibitem{dittrich2012menloreport}
{\sc Dittrich, D., and Kenneally, E.}
\newblock The menlo report: Ethical principles guiding information and
  communication technology research.
\newblock Tech. rep., 2012.

\bibitem{DML2018}
{\sc DML}.
\newblock Decentralized machine learning.
\newblock \url{https://decentralizedml.com}, 2018.

\bibitem{dwork2014algorithmic}
{\sc Dwork, C., Roth, A., et~al.}
\newblock The algorithmic foundations of differential privacy.
\newblock {\em Foundations and Trends in Theoretical Computer Science 9}, 3-4
  (2014), 211--407.

\bibitem{geyer2017FL-DP-central}
{\sc Geyer, R.~C., Klein, T., and Nabi, M.}
\newblock Differentially private federated learning: A client level
  perspective.
\newblock In {\em NIPS Workshop: Machine Learning on the Phone and other
  Consumer Devices\/} (2017).

\bibitem{googlecloud-mlaas2020}
{\sc Google}.
\newblock Ai platform.
\newblock \url{https://cloud.google.com/ai-platform}, 2020.

\bibitem{he2020fedml}
{\sc He, C., Li, S., So, J., Zeng, X., Zhang, M., Wang, H., Wang, X.,
  Vepakomma, P., Singh, A., Qiu, H., et~al.}
\newblock {FedML: A research library and benchmark for federated machine
  learning}.
\newblock {\em arXiv preprint arXiv:2007.13518\/} (2020).

\bibitem{hitaj2017deep}
{\sc Hitaj, B., Ateniese, G., and Perez-Cruz, F.}
\newblock Deep models under the gan: information leakage from collaborative
  deep learning.
\newblock In {\em Proceedings of the 2017 ACM SIGSAC Conference on Computer and
  Communications Security\/} (2017), pp.~603--618.

\bibitem{huba2022papaya-mlsys}
{\sc Huba, D., Nguyen, J., Malik, K., Zhu, R., Rabbat, M., Yousefpour, A., Wu,
  C.-J., Zhan, H., Ustinov, P., Srinivas, H., Wang, K., Shoumikhin, A., Min,
  J., and Malek, M.}
\newblock Papaya: Practical, private, and scalable federated learning.
\newblock In {\em Proceedings of Machine Learning and Systems\/} (2022),
  D.~Marculescu, Y.~Chi, and C.~Wu, Eds., vol.~4, pp.~814--832.

\bibitem{APNs}
{\sc Inc., A.}
\newblock Setting up a remote notification server.
\newblock
  \url{https://developer.apple.com/documentation/usernotifications/setting_up_a_remote_notification_server},
  2021.

\bibitem{FCM}
{\sc Inc., G.}
\newblock Firebase cloud messaging.
\newblock \url{https://firebase.google.com/docs/cloud-messaging}, 2021.

\bibitem{konecny2016federated-learning}
{\sc Konecny, J., McMahan, H.~B., Yu, F., Richtarik, P., Theertha~Suresh, A.,
  and Bacon, D.}
\newblock Federated learning: Strategies for improving communication
  efficiency.
\newblock In {\em 29th Conference on Neural Information Processing
  Systems(NIPS)\/} (2016).

\bibitem{flaas}
{\sc Kourtellis, N., Katevas, K., and Perino, D.}
\newblock Flaas: Federated learning as a service.
\newblock In {\em Proceedings of the 1st Workshop on Distributed Machine
  Learning\/} (New York, NY, USA, 2020), DistributedML'20, Association for
  Computing Machinery, p.~7–13.

\bibitem{cifar10}
{\sc Krizhevsky, A., Hinton, G., et~al.}
\newblock Learning multiple layers of features from tiny images.

\bibitem{hermes-mobicom}
{\sc Li, A., Sun, J., Li, P., Pu, Y., Li, H., and Chen, Y.}
\newblock Hermes: An efficient federated learning framework for heterogeneous
  mobile clients.
\newblock In {\em Proceedings of the 27th Annual International Conference on
  Mobile Computing and Networking\/} (New York, NY, USA, 2021), MobiCom '21,
  Association for Computing Machinery, p.~420–437.

\bibitem{marfoq2021multi-taskFL}
{\sc Marfoq, O., Neglia, G., Bellet, A., Kameni, L., and Vidal, R.}
\newblock Federated multi-task learning under a mixture of distributions.
\newblock In {\em Advances in Neural Information Processing Systems\/} (2021),
  M.~Ranzato, A.~Beygelzimer, Y.~Dauphin, P.~Liang, and J.~W. Vaughan, Eds.,
  vol.~34, Curran Associates, Inc., pp.~15434--15447.

\bibitem{flower2021}
{\sc Mathur, A., Beutel, D.~J., Porto Buarque~de Gusmao, P., Fernandez-Marques,
  J., Topal, T., Qiu, X., Parcollet, T., Gao, Y., and D.~Lane, N.}
\newblock On-device federated learning with {FLOWER}.
\newblock In {\em On-device Intelligence Workshop at the 4th Conference on
  Machine Learning and Systems (MLSys)\/} (April 2021).

\bibitem{mcmahan2017-FL}
{\sc McMahan, H.~B., Moore, E., Ramage, D., and Ag{\"{u}}era~y Arcas, B.}
\newblock Communication-efficient learning of deep networks from decentralized
  data.
\newblock In {\em 20th International Conference on Artificial Intelligence and
  Statistics (AISTATS)\/} (2017).

\bibitem{melis2019exploiting}
{\sc Melis, L., Song, C., De~Cristofaro, E., and Shmatikov, V.}
\newblock Exploiting unintended feature leakage in collaborative learning.
\newblock In {\em 2019 IEEE Symposium on Security and Privacy (SP)\/} (2019),
  IEEE, pp.~691--706.

\bibitem{azure-mlaas2020}
{\sc Microsoft}.
\newblock Azure machine learning.
\newblock \url{https://azure.microsoft.com/en-us/services/machine-learning/},
  2020.

\bibitem{ppfl}
{\sc Mo, F., Haddadi, H., Katevas, K., Marin, E., Perino, D., and Kourtellis,
  N.}
\newblock Ppfl: Privacy-preserving federated learning with trusted execution
  environments.
\newblock In {\em Proceedings of the 19th Annual International Conference on
  Mobile Systems, Applications, and Services\/} (New York, NY, USA, 2021),
  MobiSys '21, Association for Computing Machinery, p.~94–108.

\bibitem{msoon}
{\sc {Monsoon Solutions Inc.}}
\newblock {Monsoon Solutions Inc.}
\newblock \url{https://www.msoon.com}, 2022 (accessed November 29, 2022).

\bibitem{nasr2019comprehensive}
{\sc Nasr, M., Shokri, R., and Houmansadr, A.}
\newblock Comprehensive privacy analysis of deep learning: Passive and active
  white-box inference attacks against centralized and federated learning.
\newblock In {\em 2019 IEEE Symposium on Security and Privacy (SP)\/} (2019),
  IEEE, pp.~739--753.

\bibitem{openmined2020}
{\sc OpenMined}.
\newblock Openmined.
\newblock \url{https://www.openmined.org}, 2020.

\bibitem{apple-fl}
{\sc Paulik, M., Seigel, M., Mason, H., Telaar, D., Kluivers, J., van Dalen,
  R., Lau, C.~W., Carlson, L., Granqvist, F., Vandevelde, C., Agarwal, S.,
  Freudiger, J., Byde, A., Bhowmick, A., Kapoor, G., Beaumont, S., Áine
  Cahill, Hughes, D., Javidbakht, O., Dong, F., Rishi, R., and Hung, S.}
\newblock Federated evaluation and tuning for on-device personalization: System
  design \& applications, 2021.

\bibitem{rivers2014ethicalresearchstandards}
{\sc Rivers, C.~M., and Lewis, B.~L.}
\newblock Ethical research standards in a world of big.
\newblock {\em F1000Research 3\/} (2014).

\bibitem{sablayrolles2019white}
{\sc Sablayrolles, A., Douze, M., Schmid, C., Ollivier, Y., and J{\'e}gou, H.}
\newblock White-box vs black-box: Bayes optimal strategies for membership
  inference.
\newblock In {\em International Conference on Machine Learning\/} (2019),
  pp.~5558--5567.

\bibitem{s2018mobilenetv2}
{\sc Sandler, M., Howard, A., Zhu, M., Zhmoginov, A., and Chen, L.-C.}
\newblock Mobilenetv2: Inverted residuals and linear bottlenecks, 2018.

\bibitem{shokri2017membership}
{\sc Shokri, R., Stronati, M., Song, C., and Shmatikov, V.}
\newblock Membership inference attacks against machine learning models.
\newblock In {\em 2017 IEEE Symposium on Security and Privacy (SP)\/} (2017),
  IEEE, pp.~3--18.

\bibitem{tensorflow-federated2020}
{\sc TensorFlow}.
\newblock Tensorflow federated: Machine learning on decentralized data.
\newblock \url{https://www.tensorflow.org/federated}, 2020.

\bibitem{tflite-roadmap2021}
{\sc TensorFlow, G.}
\newblock Tensorflow lite roadmap.
\newblock \url{https://www.tensorflow.org/lite/guide/roadmap}, 2021.

\bibitem{tflite2022}
{\sc TensorFlow, G.}
\newblock Deploy machine learning models on mobile and iot devices.
\newblock \url{https://www.tensorflow.org/lite}, 2022.

\bibitem{batterylab}
{\sc Varvello, M., Katevas, K., Plesa, M., Haddadi, H., and Livshits, B.}
\newblock Batterylab, a distributed power monitoring platform for mobile
  devices.
\newblock In {\em Proceedings of the 18th ACM Workshop on Hot Topics in
  Networks\/} (2019), pp.~101--108.

\bibitem{yassein2017symmetric}
{\sc Yassein, M.~B., Aljawarneh, S., Qawasmeh, E., Mardini, W., and Khamayseh,
  Y.}
\newblock Comprehensive study of symmetric key and asymmetric key encryption
  algorithms.
\newblock In {\em IEEE International Conference on Engineering and Technology
  (ICET)\/} (2017).

\end{thebibliography}
